\documentclass[twoside,11pt]{article}

\usepackage[abbrvbib]{jmlr2e}
\usepackage{hyperref}
\usepackage{amsmath}             
\usepackage{algpseudocode}
\usepackage{color}
\usepackage{float}
\newcommand{\pkg}[1]{{\fontseries{b}\selectfont #1}} 
\usepackage{amssymb}
\usepackage{bm}
\usepackage{dsfont}
\usepackage[ruled,vlined]{algorithm2e}
\usepackage{footnote}

\newcommand{\ie}{\textit{i.e.}\ }
\newcommand{\eg}{\textit{e.g.}\ }

\newcommand{\etc}{etc.\ }
\newcommand{\vs}{vs.\ }

\newcommand{\alref}[1]{Algorithm~\ref{#1}}
\newcommand{\aref}[1]{Appendix~\ref{#1}}
\newcommand{\eref}[1]{Eq.~(\ref{#1})}
\newcommand{\fref}[1]{Figure~\ref{#1}}

\newcommand{\sfref}[1]{Supplementary Figure~\ref{#1}}

\newcommand{\sref}[1]{Section~\ref{#1}}

\newcommand{\stref}[1]{Supplementary Table~\ref{#1}}

\newcommand{\tref}[1]{Table~\ref{#1}}

\DeclareMathOperator{\Tr}{Tr}

\DeclareMathOperator*{\argmax}{arg\,max}

\DeclareMathOperator{\diag}{diag}
\DeclareMathOperator{\TMN}{TMN}
\DeclareMathOperator{\pd}{\partial}
\DeclareMathOperator{\dir}{Dir}
\DeclareMathOperator{\TP}{TP}
\DeclareMathOperator{\FP}{FP}
\DeclareMathOperator{\TPR}{TPR}
\DeclareMathOperator{\FPRp}{FPRp}

\ShortHeadings{Learning Bayesian Networks from Ordinal Data}{Luo, Moffa, and Kuipers}

\firstpageno{1}

\begin{document}
\title{Learning Bayesian Networks from Ordinal Data\thanks{\pkg{R} code is available at \url{https://github.com/xgluo/OSEM}}}

\author{\name Xiang Ge Luo \email xiangge.luo@bsse.ethz.ch
\\ \addr D-BSSE, ETH Zurich, Mattenstrasse 26, 4058 Basel, Switzerland
\AND
\name Giusi Moffa \\ \addr Department of Mathematics and Computer Science, University of Basel, Basel, Switzerland \\
Division of Psychiatry, University College London, London, UK 
\AND
\name Jack Kuipers \email jack.kuipers@bsse.ethz.ch
\\ \addr D-BSSE, ETH Zurich, Mattenstrasse 26, 4058 Basel, Switzerland}

\editor{\hspace{-1.5cm}{\color{white}\rule[-0.2cm]{2cm}{0.6cm}} \vspace{-1.8cm}}

\maketitle

\begin{abstract}%
Bayesian networks are a powerful framework for studying the dependency structure of variables in a complex system. The problem of learning Bayesian networks is tightly associated with the given data type. Ordinal data, such as stages of cancer, rating scale survey questions, and letter grades for exams, are ubiquitous in applied research. However, existing solutions are mainly for continuous and nominal data. In this work, we propose an iterative score-and-search method - called the Ordinal Structural EM (OSEM) algorithm - for learning Bayesian networks from ordinal data. Unlike traditional approaches designed for nominal data, we explicitly respect the ordering amongst the categories. More precisely, we assume that the ordinal variables originate from marginally discretizing a set of Gaussian variables, whose structural dependence in the latent space follows a directed acyclic graph. Then, we adopt the Structural EM algorithm and derive closed-form scoring functions for efficient graph searching. Through simulation studies, we illustrate the superior performance of the OSEM algorithm compared to the alternatives and analyze various factors that may influence the learning accuracy. Finally, we demonstrate the practicality of our method with a real-world application on psychological survey data from 408 patients with co-morbid symptoms of obsessive-compulsive disorder and depression.
\end{abstract}


\section{Introduction}

Many problems in applied statistics involve characterizing the relationships amongst a set of random variables in a complex system, aiming to describe high dimensional joint probability distributions, which may be of use in prediction and causal inference. Probabilistic graphical models, which incorporate graphical structures into probabilistic reasoning, are popular and powerful frameworks for analyzing these complex systems. The idea is to factorize the joint probability distribution $p$ for the variables $\mathbf{X} = (X_1, \cdots, X_n)^{\top}$ with respect to a graph $\mathcal{G} = (\mathcal{V},\mathcal{E})$ , where $\mathcal{V}$ is the set of nodes representing the variables, and $\mathcal{E}$ the edges encoding a set of independence relationships \citep{lauritzen1996graphical, koller2009probabilistic}. 

In this work, we focus on Bayesian networks - a special family of probabilistic graphical models where the underlying structure $\mathcal{G}$ is a directed acyclic graph (DAG) - also named DAG models in the literature \citep{geiger2002parameter, pearl2014probabilistic}. The joint probability distribution $p$ can be fully specified by a set of parameters $\theta$ and factorizes according to $\mathcal{G}$ as
\begin{equation}
\label{eq:factorization}
p(\mathbf{x} \mid \theta, \mathcal{G}) = p(x_1, \dots, x_n \mid \theta, \mathcal{G}) = \prod^n_{i = 1} p(x_i \mid \mathbf{x}_{pa(i)}, \theta_i, \mathcal{G}),
\end{equation}
where $\mathbf{x}$ is a realization for $\mathbf{X}$, $\theta = \cup^n_{i = 1} \theta_i$, and the subsets $\{\theta_i\}_{i = 1}^n$ are assumed to be disjoint. We denote the parents of node $i$ by $pa(i)$ such that there is an directed edge from $j$ to $i$ if $j \in pa(i)$. Hence, we can also read the factorization in \eqref{eq:factorization} by saying that a variable $X_i$ is conditionally independent from its non-descendants given its parents $\mathbf{X}_{pa(i)}$ in $\mathcal{G}$. This is known as the Markov property \citep{lauritzen1996graphical}. We denote a Bayesian network by a set $\mathcal{B} = (\mathcal{G},\theta)$. Given a data sample $\mathcal{D}$, learning a Bayesian network, therefore, means estimating both the network structure $\mathcal{G}$ and the parameters $\theta$. 

\subsection{Structure Learning for Bayesian Networks}
\label{sec:intro_struct}

Structure learning for Bayesian networks is an NP-hard problem \citep{chickering2004large}, mainly because the number of possible DAGs grows super-exponentially with the number of nodes $n$. Existing approaches to tackle this problem fall roughly into three categories: 
\begin{itemize}
\item Constraint-based methods perform conditional independence tests for each pair of nodes given a subset of adjacent nodes in order of increasing complexity. Examples are the PC \citep{spirtes2000causation}, FCI \citep{spirtes2001anytime}, and RFCI algorithms \citep{colombo2012learning}. These methods are fast and computationally feasible when the underlying graph is sparse \citep{kalisch2007estimating,uhler2013geometry}.

\item Score-and-search methods rely on a scoring function and an algorithm that searches through the DAG space. One example is the greedy equivalence search (GES), which starts with an empty graph and goes through two phases of adding and removing edges based on the scores \citep{chickering2002optimal}. Recently, score-and-search algorithms based on integer linear programming have also been introduced \citep{cussens2017bayesian}. More general approaches to score-and-search include sampling based methods, which aims to construct the full posterior distribution of DAGs given the data \citep{madigan1995bayesian, giudici2003improving, friedman2003being, grzegorczyk2008improving, kuipers2017partition}, as well as strategies relying on dynamic programming \citep{koivisto2004exact}. These methods can be slower than the constraint-based ones, but in general, provide better performance when computationally feasible \citep{heckerman1999bayesian}.

\item Lastly, the hybrid approaches bring together the above two solutions by restricting the initial search space using a constraint-based method in order to improve the speed and accuracy even further \citep{tsamardinos2006max,nandy2018high,kuipers2018efficient}. 
\end{itemize}
For a broader overview on Bayesian networks we refer to the comprehensive review by \cite{daly2011learning} as well as the textbook by \cite{koller2009probabilistic}. A recent comparison of structure learning approaches can be found in \cite{constantinou2021large} for example.

\subsection{Structure Learning with Ordinal Data}
\label{sec:intro_ordinal}

Alongside the search scheme, it is also important to properly account for the type of random variables studied and define a suitable scoring function. The focus of this work is on categorical variables, which can be either nominal or ordinal, depending on whether the ordering of the levels is relevant. Examples of nominal variables include sex (male, female, non-binary), genotype (AA, Aa, aa), and fasting before a blood test (yes, no). They are invariant to any random permutation of the categories. The listing order of the categories is, by contrast, an inherent property for ordinal variables. Examples are stages of disease (I, II, III), survey questions with Likert scales (strongly disagree, disagree, undecided, agree, strongly agree), as well as discretized continuous data, such as the body mass index (underweight, normal weight, overweight, obese), age groups (children, youth, adults, seniors) \etc \citep{mcdonald2009handbook, agresti2010analysis}. For discretized data, it is often difficult or impossible to access the underlying continuous source for practical or confidential reasons. The interactions amongst the latent variables can only manifest through their ordinal counterparts. From a technical perspective it is thus natural to think of ordinal variables as generated from a set of continuous latent variables through discretization.

Despite the prevalence of ordinal variables in applied research, not much attention has been paid to the problem of learning Bayesian networks from ordinal data. One solution is to apply constraint-based methods with a suitable conditional independence test, including but not limited to the Jonckheere-Terpstra test in the OPC algorithm of \cite{musella2013pc}, copula-based tests \citep{cui2016copula}, and likelihood-ratio tests \citep{tsagris2018constraint}. While a possible extension along this line is to explore more appropriate tests to improve the search, constraint-based methods, in general, tend to favor sparser graphs and can be dependent on the testing order \citep{uhler2013geometry,colombo2014order}. In the score-and-search regime, no scoring functions exist for ordinal variables, so metrics ignoring the ordering and based on the multinomial distribution are the typical alternatives. An obvious drawback is the loss of information associated with ignoring the ordering among the categories, resulting in inaccurate statistical analyses. Another potential issue is the tendency to overparameterization, especially when the number of levels is greater than $3$. In this case, neither statistical nor computational efficiency can be achieved \citep{talvitie2019learning}. However, for very large number of levels a continuous approximation may be adequate.

In this work, we develop an iterative score-and-search scheme specifically for ordinal data. Before describing the algorithm in detail, we briefly review the challenges. On the one hand, we need a scoring function that can preserve the ordinality amongst the categories. On the other hand, that function should also satisfy three important properties: decomposability, score equivalence, and consistency \citep{koller2009probabilistic}. Decomposability of a score is the key to fast searching. Modifying a local structure does not require recomputing the score of the entire graph, which tremendously reduces the computational overhead. Score equivalence is less crucial but is necessary for search schemes relying on Markov equivalent classes, such as GES or hybrid methods that initialize with a PC output. Here, two DAGs belong to the same Markov equivalent class if and only if they share the same skeleton and v-structures \citep{verma1991equivalence}. A Markov equivalent class can be uniquely represented by a completed partially directed acyclic graph (CPDAG). Finally, a consistent score can identify the true DAG if sufficient data is provided. Examples of scores that satisfy all three properties are the BDe score \citep{heckerman1995learning} for (nominal) categorical data and the BGe score \citep{geiger2002parameter,kuipers2014addendum} for continuous data. It is, however, nontrivial to develop such a score for ordinal data. 

\begin{figure}
\begin{center}
\includegraphics[width = \textwidth, height = 5cm]{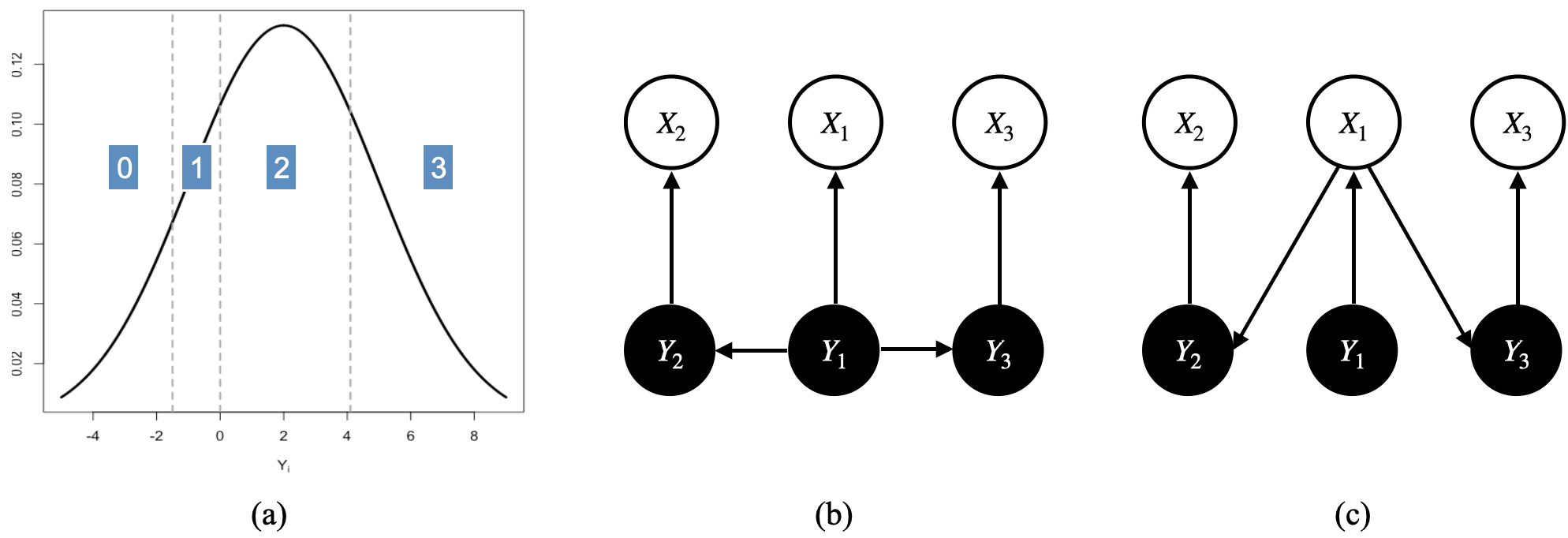}
\caption[Illustration of the probit models]{(a) We assume that an ordinal variable is obtained by marginally discretizing a latent Gaussian variable. (b) and (c) are examples of three-node graphs for two different probit models discussed in \sref{sec:intro_ordinal}. $X_1$, $X_2$ and $X_3$ are ordinal variables, each obtained by discretizing a latent variable $Y_i$ with associated Gaussian parameters $\theta_i$, $i = 1,2,3$. Latent nodes are shaded for clarity.}
\label{fig:Type_I_II}
\end{center}
\end{figure}

To capture the ordinality, we use a widely accepted parameterization based on the multivariate probit models \citep{ashford1970multi,bock1996high,chib1998analysis, daganzo2014multinomial}, with no additional covariates with respect to the variables whose joint distribution we wish to model and hence focusing on their dependency structure. As illustrated in \fref{fig:Type_I_II}(a), each ordinal variable is assumed to be obtained by marginally discretizing a latent Gaussian variable. In this case, the order of the categories is encoded by the continuous nature of the Gaussian distribution.

More specifically we can have at least two possible formulations, both with their pros and cons. The first one (\fref{fig:Type_I_II}(b)) assumes that the latent variables $\mathbf{Y}$ jointly follow a multivariate Gaussian distribution, which factorizes according to a DAG. This approach resembles the Gaussian DAG model \citep{heckerman1995learning, geiger2002parameter} in the latent space, with additional connections to the ordinal variables $\mathbf{X}$. Unlike in the continuous case, marginalizing out the latent variables will result in a non-decomposable observed likelihood. The conditional independence relationships will also disappear. This is because DAG models are not closed under marginalization and conditioning \citep{richardson2002ancestral, silva2009hidden, colombo2012learning}. In \fref{fig:Type_I_II}(b), the Gaussian variables $Y_2$ and $Y_3$ are conditionally independent given $Y_1$, whereas the ordinal variables $X_2$ and $X_3$ are not independent given $X_1$ due to the presence of a common ancestor $Y_1$ in the latent space. As a result, traditional closed-form scoring functions such as the BGe metric \citep{heckerman1995learning} cannot be applied, which is generally the case in the presence of hidden variables.

Another setup (\fref{fig:Type_I_II}(c)) uses a non-linear structural equation model with the probit function as the link. In this case it is assumed that a DAG structure describes the relationship between the observed variables, whereas the latent Gaussian variables only facilitate the flow of information and can be marginalized out. It is equivalent to the probit directed mixed graph model of \cite{silva2009hidden}. In this model, the observed likelihood is decomposable. We can recover the graph $X_2 \leftarrow X_1 \rightarrow X_3$ after marginalization. However, the resulting score is not score-equivalent because the joint probability distribution fails to meet the complete model equivalence assumption as described by \cite{geiger2002parameter}.

Our solution builds on the first formulation, which we call the \textit{latent Gaussian DAG model}. The key is to observe that the complete-data likelihood is both decomposable and score-equivalent. Further adding a BIC-type penalty will still preserve these key properties of the score, as well as consistency \citep{schwarz1978estimating, rissanen1987stochastic, barron1998minimum}. The Structural EM algorithm of \cite{friedman1997learning} addresses the problem of structure learning in the presence of incomplete data by embedding the search for the optimal structure into an Expectation-Maximization (EM) algorithm. We apply the Structural EM to deal with the latent Gaussian variable construction of ordinal data in a procedure which we call the \textit{Ordinal Structural EM (OSEM)} algorithm. Note that \cite{webb2008bayesian} also used the latent Gaussian DAG model and aimed to determine the network structure from the ordinal data using a reversible jump MCMC method. Our approach differs in that it explicitly exploits score decomposability, allowing for search schemes that are computationally more efficient, such as order and partition MCMC \citep{friedman2003being, kuipers2017partition}, or hybrid methods.

The rest of this article is structured as follows. In \sref{sec:model}, we formally define the latent Gaussian DAG model and introduce the corresponding identifiability constraints. In \sref{sec:OSEM}, we integrate the model into the Structural EM framework. In \sref{sec:experiments}, we use synthetic and real data to illustrate the superior performance of our proposed method compared to the alternatives in terms of structure recovery and prediction. In \sref{sec:psych}, we apply our method to a psychological survey data set. Finally, in \sref{sec:discussion}, we discuss the implications of our results as well as possible directions for future work.

\section{The Latent Gaussian DAG Model}
\label{sec:model}

Let $\mathbf{X} = (X_1, \cdots, X_n)^{\top}$ be a collection of $n$ ordinal variables, where $X_i$ takes values in the set $\{\tau(i, 1), \tau(i, 2), \dots, \tau(i, L_i)\}$ with $\tau(i, 1) < \tau(i, 2) < \dots < \tau(i, L_i)$, $i = 1,\dots,n$. We assume that the number of levels $L_i \geq 2$, so each variable should at least be binary. It is common to set $\tau(i, j) = j - 1$ for all $1 \leq j \leq L_i$, \ie $\tau(i, 1) = 0$, $\tau(i, 2) = 1$, and so on. Further, we assume that each $X_i$ is obtained by discretizing an underlying Gaussian variable $Y_i$ using the thresholds $-\infty =: \alpha(i,0) < \alpha(i,1) < \dots < \alpha(i, L_i - 1) < \alpha(i,L_i) := \infty$. Let $\bm{\alpha}_i = (\alpha(i,0), \dots, \alpha(i,L_i))^{\top}$ and $\bm{\alpha} = \{\bm{\alpha}_i\}_{i=1}^n$. The hidden variables $\mathbf{Y} = (Y_1, \cdots, Y_n)^{\top}$ jointly follow a multivariate Gaussian distribution $\mathcal{N}(\bm{\mu}, \Sigma)$, which factorizes according to some DAG $\mathcal{G}$. Formally, the latent Gaussian DAG model is given by
\begin{equation}
\label{eq:model_ordinal}
\begin{split}
& Y_i \mid \mathbf{y}_{pa(i)}, \vartheta_i, \mathcal{G} \sim \mathcal{N}(\mu_i + \sum_{j \in pa(i)} b_{ji} (y_j - \mu_j), v_i)\\
& P(X_i = \tau(i,l) \mid Y_i = y_i, \bm{\alpha}_i) =  \mathds{1} \big( y_i \in [\alpha(i,l-1), \alpha(i,l)) \big), \quad l = 1,\dots, L_i \\
& p(\mathbf{x}, \mathbf{y} \mid \theta, \mathcal{G}) = \prod^n_{i = 1} \phi(y_i \mid \mathbf{y}_{pa(i)}, \vartheta_i, \mathcal{G})p(x_i \mid y_i, \bm{\alpha}_i)
\end{split}
\end{equation}
where $\theta = \cup^n_{i = 1}\theta_i$ with $\theta_i = (\vartheta_i, \bm{\alpha}_i)$, $\vartheta_i = (\mu_i, \mathbf{b}_i, v_i)$ and $\mathbf{b}_i = (b_{ji})_{j \in pa(i)}$ for all $i = 1,\dots,n$. Since the discretization is marginal and deterministic, the joint probability distribution $p(\mathbf{x}, \mathbf{y} \mid \theta, \mathcal{G})$ remains decomposable and score-equivalent as in the Gaussian DAG model \citep{geiger2002parameter}. Alternatively we can parametrize the complete joint distribution of the hidden and observed variables using a mean vector $\bm{\mu} = (\mu_1, \dots, \mu_n)^{\top} \in \mathbb{R}^n$ and a symmetric positive definite covariance matrix $\Sigma \in \mathbb{R}^{n\times n}$,
\begin{equation}
\label{eq:reparam}
p(\mathbf{x}, \mathbf{y} \mid \theta, \mathcal{G}) = \phi(\mathbf{y} \mid \bm{\mu}, \Sigma, \mathcal{G}) p(\mathbf{x} \mid \mathbf{y}, \bm{\alpha}).
\end{equation} 
It follows from \cite{silva2009hidden} that we can write the transformation between $\{\mathbf{b}_i, v_i\}_{i = 1}^n$ and $\Sigma$ as
\begin{equation}
\label{eq:Sigma_transform}
\Sigma = (\mathbf{I} - \mathbf{B})^{-1} \mathbf{V} (\mathbf{I} - \mathbf{B})^{- \top},
\end{equation}
where $(\mathbf{B})_{ij} = b_{ji}$ and $\mathbf{V}$ is an $n$-by-$n$ diagonal matrix with $\mathbf{V}_{ii} = v_i$.

\subsection{Identifiability}
\label{sec:identifiability}

Different underlying Gaussian variables $\mathbf{Y}$ may generate the same contingency table for $\mathbf{X}$, by simply shifting and scaling the thresholds using the corresponding means and variances. For example, we can obtain the same ordinal variable $X$ by discretizing either $Y_1 \sim \mathcal{N}(0,1)$ at $\{-1,1\}$ or $Y_2 \sim \mathcal{N}(-1,100)$ at $\{-11,9\}$. In other words, there is not a one-to-one mapping between the cell probabilities in the table and the parameters $\theta = (\bm{\alpha}, \bm{\mu}, \Sigma)$. Thus, we need to impose some constraints to ensure model identifiability. 

Typically, each dimension of $\mathbf{X}$ requires two constraints.  \cite{webb2008bayesian} choose to fix the lowest and the highest thresholds for deriving the ordinal variables. For binary variables, they restrict the only threshold at zero and set $v_i = 1$. Instead, we find it computationally more convenient to standardize each latent dimension. More precisely, we let $\mu_i = 0$ for all $i = 1,\dots,n$ and constrain the covariance matrix to be in  correlation form using a diagonal matrix $D := \diag(d_1, \dots, d_n)$ with $d_i = \sqrt{\Sigma_{ii}}$, \ie we replace $\Sigma$ by its transformation $D^{-1}\Sigma D^{-1}$. Due to its symmetry the correlation matrix is identified by $\frac{n(n-1)}{2}$ off-diagonal parameters $\Sigma_{ij} = \Sigma_{ji} = \rho_{ij}$ for all $i \neq j$. Imposing these constraints ensures that the thresholds and the correlation matrix will be identifiable, which is sufficient for deducing the hidden DAG $\mathcal{G}$. The mean vector and the variances remain unidentifiable.

\section{The Ordinal Structural EM Algorithm}
\label{sec:OSEM}

Given a sample $\mathcal{D}_{\mathbf{X}} = \{\mathbf{x}^1, \dots, \mathbf{x}^N\}$ of size $N$ for $\mathbf{X}$, our goal is to learn both the parameters and the structure of the Bayesian network $\mathcal{B} = (\mathcal{G},\theta)$ which best explains the observed data. Making inference about Bayesian networks with unknown structure is an intrinsically challenging problem, and in the case of ordinal variables an additional difficulty originates from the presence of latent variables. To estimate the parameters $\theta$ for a given DAG $\mathcal{G}$, we can use the Expectation-Maximization (EM) algorithm \citep{dempster1977maximum}, a common approach to handle maximum likelihood estimation in the presence of missing data. Conversely, with fully observed data, we may use one of the methods discussed in \sref{sec:intro_struct} for structure learning. In the presence of hidden variables, the above strategies fail, because the marginal density for $\mathbf{X}$ no longer decomposes over each node and its parents. In the sequel, we consider combining the latent Gaussian DAG model with the Structural EM algorithm of \cite{friedman1997learning}, where the structure learning step is wrapped inside the EM procedure. The resulting framework is the Ordinal Structural EM (OSEM) algorithm.

\subsection{The EM and the Structural EM Algorithms}
\label{sec:EM}

In the presence of latent variables, computing the maximum likelihood estimates for $\theta$ by maximizing the observed data log-likelihood 
\begin{equation}
\label{eq:observedLL}
\ell(\theta; \mathcal{D}_{\mathbf{X}}) = \log p(\mathcal{D}_{\mathbf{X}} \mid \theta) = \log \int p(\mathcal{D}_{\mathbf{X}}, \mathcal{D}_{\mathbf{Y}} \mid \theta) d\mathbf{y}^1 \cdots d \mathbf{y}^N
\end{equation}
can be intractable. The EM algorithm addresses this problem by alternating between two steps:
\begin{itemize}
\item \textbf{E-step} (Expectation): given the observed data and the current estimate of the parameters $\theta^{(t)}$, $t \in \{0,1,2,\dots\}$, compute the expected value of the complete-data (including the hidden variables) likelihood with respect to the distribution of the hidden variables conditional on the observed variables,
\begin{equation}
\label{eq:Q_function}
\begin{split}
Q(\theta,\theta^{(t)}) & = \mathbb{E}_{\mathbf{Y} \mid \mathbf{x}, \theta^{(t)}}[\log p(\mathcal{D}_{\mathbf{X}}, \mathcal{D}_{\mathbf{Y}} \mid \theta)] \\
& = \sum_{j = 1}^N \mathbb{E}_{\mathbf{Y}^j \mid \mathbf{x}^j, \theta^{(t)}}[\log p(\mathbf{X}^j, \mathbf{Y}^j \mid \theta) \mid \mathbf{X}^j = \mathbf{x}^j];
\end{split}
\end{equation}
\item \textbf{M-step} (Maximization): update the parameters by maximising the $Q$ function with respect to $\theta$ as
\begin{equation}
\label{eq:Q_max}
\theta^{(t+1)} = \argmax_{\theta \in \Theta} Q(\theta,\theta^{(t)}).
\end{equation}
\end{itemize}
The EM procedure guarantees convergence of the observed data likelihood $\ell(\theta^{(t)}; \mathcal{D}_{\mathbf{X}})$ to a local maximum as $t \rightarrow \infty$ \citep{dempster1977maximum}. 

The Structural EM algorithm is an extension of the original EM for learning Bayesian networks. The idea is to recursively choose a structure and a set of parameters that improve the following \textit{expected scoring function}:
\begin{equation}
\label{eq:SEM_Q}
\tilde{Q}(\mathcal{G}, \theta; \mathcal{G}^{(t)}, \theta^{(t)}) = \underbrace{\mathbb{E}_{\mathbf{Y} \mid \mathbf{x},\mathcal{G}^{(t)}, \theta^{(t)}}[\log p(\mathcal{D}_{\mathbf{X}}, \mathcal{D}_{\mathbf{Y}} \mid \mathcal{G}, \theta)]}_{Q(\mathcal{G}, \theta; \mathcal{G}^{(t)}, \theta^{(t)})} - \text{Penalty}(\mathcal{G}, \theta).
\end{equation}
Here, we choose a BIC penalty term in order to ensure that the score \eref{eq:SEM_Q} remains decomposable, score-equivalent, and consistent (Theorem 18.2 of \cite{koller2009probabilistic}). In the following sections, we elaborate on the steps in the OSEM algorithm.

\subsection{Initialization}
\label{sec:OSEM_init}

The original algorithm of \cite{friedman1997learning} used a random initialization for both the structure $\mathcal{G}$ and the parameters $\theta$. However, we can achieve faster convergence if we can cheaply obtain rough estimates of the parameters which maximise the penalised likelihood function, and start the EM procedure from such estimates.

Given the identifiability constraints in \sref{sec:identifiability}, we only need to estimate the thresholds $\bm{\alpha}$ once at the very beginning. More precisely, given a sample $\mathcal{D}_{\mathbf{X}}$, we can estimate the thresholds $\alpha(i,l)$ for all $i = 1,\dots,n$ and $l = 1,\dots, L_i - 1$ as 
\begin{equation}
\label{eq:threshold}
\hat{\alpha}(i,l) = \Phi^{-1} \Big( \frac{1}{N} \sum^N_{j = 1} \mathds{1}(x^j_i \leq \tau(i,l)) \Big),
\end{equation}
which are the empirical quantiles of the standard normal distribution.

Next, we assume that the initial DAG $\mathcal{G}^{(0)}$ is a full DAG. Conditioned on the thresholds, we want to find the correlation matrix by maximizing the observed data log-likelihood, 
\begin{equation}
\Sigma^{(0)} \mid \hat{\bm{\alpha}} = \argmax_{\Sigma} \log p(\mathcal{D}_{\mathbf{X}} \mid \mathbf{0},\Sigma, \mathcal{G}^{(0)}), 
\end{equation}
which can be computationally too expensive. Alternatively, we can use a pairwise likelihood approach \citep{kuk2000pairwise, renard2004pairwise, varin2010mixed}, where we estimate each off-diagonal entry $\rho_{ij}$ in the correlation matrix as
\begin{equation}
\label{eq:pairwise}
\rho^{(0)}_{ij} \mid \hat{\bm{\alpha}} = \argmax_{\rho_{ij}}  \log p \bigg(\mathcal{D}_{X_i \cup X_j} \mid 
\begin{bmatrix}
0 \\
0
\end{bmatrix}
,
\begin{bmatrix}
1 &  \rho_{ij}\\
\rho_{ij} & 1
\end{bmatrix}
, \mathcal{G}^{(0)} \bigg).
\end{equation}
Let $\theta^{(0)} = \{\Sigma^{(0)}, \hat{\bm{\alpha}}\}$ with $(\Sigma^{(0)})_{ij} = \rho^{(0)}_{ij}$ for all $i \neq j$ and $(\Sigma^{(0)})_{ii} = 1$ for all $i = 1,\dots,n$. This approach is fast and easy to implement at the expense of accuracy. If the resulting correlation matrix is not positive-definite, one remedy is to smooth the matrix by coercing the non-positive eigenvalues into slightly positive numbers. As this is only an initialization, the difference caused by smoothing should not severely impact the following stages. 

\subsection{Structure Update}
\label{sec:OSEM_struct}

The algorithm iterates through two steps: the structure update and the parameter update. Given the current estimate $(\mathcal{G}^{(t)}, \theta^{(t)})$, the structure update searches for the DAG that maximizes the expected scoring function by computing the expected statistics instead of the actual ones in the complete-data setting.

In our case, computing the expected values in \eref{eq:Q_function} is expensive, so we use the Monte Carlo EM algorithm of \cite{wei1990monte} as an approximation. For each observation $\mathbf{x}^j$, we draw a sample of size $K$ from the truncated multivariate normal distribution
\begin{equation}
\label{eq:TMN}
\mathbf{Y}^j \mid \mathbf{x}^j, \Sigma^{(t)}, \hat{\bm{\alpha}}, \mathcal{G}^{(t)} \sim \TMN(\bm{0},\Sigma^{(t)} \mid \mathbf{x}^j, \hat{\bm{\alpha}}),
\end{equation}
where the sampling region of the Gaussian data is restricted to the domain specified by $\mathbf{x}^j$ and $\hat{\bm{\alpha}}$. Then, we can approximate the expected covariance matrix as
\begin{equation}
\label{eq:expected_Sigma}
\begin{split}
\hat{\Sigma} = \frac{1}{N} \sum^N_{j = 1} \mathbb{E}_{\mathbf{Y}^j \mid \mathbf{x}^j,\mathcal{G}^{(t)},  \theta^{(t)}} \big[(\mathbf{Y}^j) (\mathbf{Y}^j)^{\top} \big] \approx \frac{1}{N} \frac{1}{K} \sum^N_{j = 1} \sum^K_{k = 1} ( \mathbf{y}^{j(k)}) ( \mathbf{y}^{j(k)})^{\top}.
\end{split}
\end{equation}
From \eref{eq:model_ordinal} and \eref{eq:reparam}, we decompose the $Q$ function with respect to each node and its parents,
\begin{equation}
\label{eq:struct_interm}
Q(\mathcal{G}, \theta; \mathcal{G}^{(t)}, \theta^{(t)}) = \sum^n_{i = 1} \sum_{j = 1}^N \mathbb{E}_{\mathbf{Y}^j \mid \mathbf{x}^j, \mathcal{G}^{(t)}, \theta^{(t)}}\bigg[\log \frac{\phi(\mathbf{Y}^j_{\{i\} \cup pa(i)} \mid (\bm{0},\Sigma)_{\{i\} \cup pa(i)}, \mathcal{G})}{ \phi(\mathbf{Y}^j_{pa(i)}\mid (\bm{0},\Sigma)_{pa(i)}, \mathcal{G})} \bigg],
\end{equation}
where $(\bm{0},\Sigma)_O$ are the parameters indexed by the set $O$ for $O \subseteq \{1,\dots,n\}$. Let $\hat{\theta} = \{\hat{\Sigma}, \hat{\bm{\alpha}}\}$. By substituting \eref{eq:expected_Sigma} into \eref{eq:struct_interm} and using the standard normal density function, we rewrite \eref{eq:SEM_Q} as
\begin{equation}
\label{eq:struct_BICScore}
\begin{split}
& \tilde{Q}(\mathcal{G}, \hat{\theta}; \mathcal{G}^{(t)}, \theta^{(t)}) = Q(\mathcal{G}, \hat{\theta} ; \mathcal{G}^{(t)}, \theta^{(t)}) - \frac{\log N}{2} \dim(\mathcal{G})\\
& =  \sum^n_{i = 1} \bigg( -\frac{N}{2} \bigg[\log \Big(\hat{\Sigma}_{i,i} - \hat{\Sigma}_{i,pa(i)} \hat{\Sigma}_{pa(i),pa(i)}^{-1} \hat{\Sigma}_{pa(i),i} \Big) \bigg] - \frac{\log N}{2} \dim(Y_i,\mathbf{Y}_{pa(i)}) \bigg),
\end{split}
\end{equation}
which is an expected version of the BIC score as in the complete-data case. Therefore, to update the structure we need to solve the following maximization problem,
\begin{equation}
\label{eq:struct_update}
\mathcal{G}^{(t+1)} = \argmax_{\mathcal{G}: \text{ DAG over } (\mathbf{X},\mathbf{Y})} \tilde{Q}(\mathcal{G}, \hat{\theta}; \mathcal{G}^{(t)}, \theta^{(t)}),
\end{equation}
which we can address by using any of the existing score-based or hybrid schemes, such as the GES and MCMC samplers.

\subsection{Parameter Update}
\label{sec:OSEM_param}

Conditioned on the structure estimate $\mathcal{G}^{(t+1)}$, the goal in the parameter update is to find $\theta^{(t+1)}$ such that
\begin{equation}
\label{eq:param_update}
\begin{split}
\theta^{(t+1)} = \argmax_{\theta \in \Theta} Q(\mathcal{G}^{(t+1)}, \theta; \mathcal{G}^{(t)}, \theta^{(t)}) - \frac{\log N}{2} \dim(Y_i,\mathbf{Y}_{pa(i)}).
\end{split}
\end{equation}
The $Q$ function can be expressed in terms of the conditional parameters from \eref{eq:model_ordinal}:
\begin{equation}
\begin{split}
& \quad \sum^n_{i = 1} \sum_{j = 1}^N \mathbb{E}_{\mathbf{Y}^j \mid \mathbf{x}^j, \mathcal{G}^{(t)}, \theta^{(t)}}\Big[\log \phi(Y^j_i \mid \mathbf{Y}^j_{pa(i)}, \theta_i, \mathcal{G}^{(t+1)}) \Big] \\
& =\sum^n_{i = 1} - \frac{N}{2} \bigg(\log(2\pi v_i)  + \frac{1}{v_i} \frac{1}{N} \sum_{j = 1}^N \underbrace{\mathbb{E}_{\mathbf{Y}^j \mid \mathbf{x}^j,\mathcal{G}^{(t)},  \theta^{(t)}} \Big[\big(Y^j_i - \mathbf{b}_i^{\top} \mathbf{Y}^j_{pa(i)}  \big)^2 \Big]}_{\approx \frac{1}{K} \sum^K_{k = 1} \Big[\big(y^{j(k)}_i - \mathbf{b}_i^{\top} \mathbf{y}^{j(k)}_{pa(i)}  \big)^2 \Big]} \bigg).
\end{split}
\end{equation}
The truncated multivariate normal samples in the previous structure update can be recycled here, because the parameters $\theta^{(t)}$ have not yet been updated. To maximize $Q(\mathcal{G}^{(t+1)}, \theta; \mathcal{G}^{(t)}, \theta^{(t)})$ we can then effectively use the  multi-step conditional maximization \citep{meng1993maximum} following the topological order in $\mathcal{G}^{(t+1)}$. Suppose we have estimated $\theta^{(t+1)}_{pa(i)}$ and consider now updating $\theta_i$. Let $\mathbf{Y}_{pa(i)}^{(k)} = (\mathbf{y}^{1(k)}_{pa(i)} ,\dots, \mathbf{y}^{N(k)}_{pa(i)} )^{\top} \in \mathbb{R}^{N \times \lvert pa(i) \rvert}$ and $\mathbf{y}^{(k)}_i = (y^{1(k)}_i, \dots, y^{N(k)}_i)^{\top} \in \mathbb{R}^N$ represent the sampled design matrix and the sampled response vector respectively. We first differentiate $Q(\mathcal{G}^{(t+1)}, \theta; \mathcal{G}^{(t)}, \theta^{(t)})$ with respect to $\mathbf{b}_i$ and obtain
\begin{equation}
\label{eq:param_b}
\mathbf{b}^{(t+1)}_i = \bigg(\sum^K_{k = 1} \Big(\mathbf{Y}_{pa(i)}^{(k)} \Big)^{\top} \mathbf{Y}_{pa(i)}^{(k)} \bigg)^{-1} \bigg(\sum^K_{k = 1} \Big(\mathbf{Y}_{pa(i)}^{(k)} \Big)^{\top} \mathbf{y}^{(k)}_i \bigg).
\end{equation}
Conditioned on $\mathbf{b}^{(t+1)}_i$, the update for $v_i$ is then
\begin{equation}
\label{eq:param_v}
v^{(t+1)}_i = \frac{1}{N} \frac{1}{K} \sum_{j = 1}^N  \sum^K_{k = 1} \Big( y^{j(k)}_i - (\mathbf{b}^{(t+1)}_i)^{\top} \mathbf{y}^{j(k)}_{pa(i)} \Big)^2.
\end{equation}
By introducing the BIC penalty, this procedure is equivalent to performing nodewise regression with best subset selection \citep{james2013introduction} in the latent space. Only minor modifications in the active components of $\mathbf{b}^{(t+1)}_i$ are required.

For the purpose of identifiability and the next structure update, the final step is to convert $\theta^{(t+1)} = (\mathbf{b}^{(t+1)}_i, v^{(t+1)}_i)_{i = 1}^n$ to its corresponding correlation form with \eref{eq:Sigma_transform} and rescaling.

\subsection{Summary}
\label{sec:OSEM_summary}

The latent Gaussian DAG model assumes that the ordinal variables originate from element-wise discretization of a set of Gaussian variables, which follow a DAG structure. Given observations from the ordinal data only, we call Ordinal Structural EM (OSEM) algorithm the process of estimating the hidden DAG structure, and we summarize the procedure in \alref{algo:OSEM}, along with more technical details (\aref{sec:algorithm}). Theorem 3.1 of \cite{friedman1997learning} states that the penalized observed data log-likelihood can only improve at each iteration, and therefore, guarantees the convergence of the algorithm to a local optimum.

\section{Experimental Results}
\label{sec:experiments}

In this section, we first use simulation studies to assess the performance of our proposed method in recovering the hidden network structure from observed ordinal data. Then, we evaluate its predictive performance in terms of average log loss using real data sets. 

\subsection{Structure Recovery}
\label{sec:structure_recovery}

We randomly generate DAGs using the \texttt{randDAG} function from the \textbf{\texttt{pcalg}} package \citep{kalisch2012causal}, followed by generating the corresponding Gaussian data according to the topological order in the DAGs. Then, we discretize the Gaussian data to obtain an ordinal data set, from which we try to learn the original structure describing the relationship between the latent Gaussian variables. We compare our Ordinal Structural EM algorithm (OSEM) to five other existing approaches: 
\begin{enumerate}

\item PC algorithm with the nominal $G^2$ test (NPC);
\item PC algorithm of \cite{musella2013pc} with the ordinal Jonckheere--Terpstra test (OPC);
\item PC algorithm with the Gaussian test using Fisher's z--transformation (GPC);
\item hybrid method with the BDe score and NPC as initial search space (BDe);
\item hybrid method with the BGe score and GPC as initial search space (BGe).

\end{enumerate}
Note that none of these methods assume a latent Gaussian DAG model. Namely, they all build a network model directly on the observed ordinal variables. OSEM on the other hand estimates the hidden structure on the latent Gaussian variables. Therefore a direct comparison of the performance with respect to the network structure needs to be taken with caution, especially in light of the fact that conditional independence relationships do not automatically extend from the latent to the observed space (see also \sref{sec:intro_ordinal}). To the best of our knowledge, however, no generative model for ordinal data currently exists that provides both decomposability and score equivalence in the observed space. In the absence of alternative methods which are capable of dealing with the latent variable construction of ordinal data, it may still be insightful to compare the network accuracy of OSEM to that of the most popular methods currently in use to learn network models for various data types. Finally, to compare OSEM to other algorithms in a manner that is agnostic to the data generating scheme, we will evaluate predictive performance on real data sets in \sref{sec:pred}. Detailed implementations for all methods are summarised in \aref{sec:r_func}. 

\subsubsection{Performance Metrics}
\label{sec:metrics}

We assess the performance of the model by comparing the estimated structure with the true structure. Notice that we can only identify the true DAG up to its Markov equivalence class, so it is more appropriate to compare the corresponding CPDAGs. However, the metrics based on CPDAG differences can still be too harsh, because many directed edges are induced by a few number of v-structures. Therefore, we choose to compare the patterns of the DAGs in the sense of \cite{meek1995causal}, where the v-structures are the only directed edges. In other words, we measure the structural hamming distance (SHD) between two DAGs using the skeleton differences plus the v-structure differences. 

The estimated DAG or CPDAG is first converted into a pattern. If an edge in the estimated pattern matches exactly the same direction as the corresponding edge in the true pattern, we count it as $1$ true positive (TP). If the direction is wrong but exactly one of this edge and the corresponding edge is undirected, then we count it as $0.5$ TP. We subtract the number of true positives from the total number of estimated edges to get the number of false positives (FP). Furthermore, we compute the true positive rate (TPR) and the false positive rate (FPRp) as
\begin{equation}
\TPR = \frac{\TP}{\text{P}} \qquad \text{and} \qquad \FPRp = \frac{\FP}{\text{P}},
\label{eq:FPRp}
\end{equation}
where P is the total number of edges in the true pattern. Scaling the false positives by P instead of the number of real negatives N can lead to a more comparable visualization, because if the sparsity of the network stays the same, N grows quadratically as $n$ increases. We illustrate the comparisons mainly through Receiver Operating Characteristics (ROC) curves. In particular, we plot the TPR against the FPRp with the following penalization parameters:
\begin{itemize}
\item significance level for the conditional independence tests:

$\alpha \in \{0.001,0.01,0.025,0.05,0.075,0.1,0.15,0.2,0.25,0.3\}$.

\item equivalent sample size for the BDe score: 

$\chi \in \{0.0001,0.001,0.01,0.1,1,10,20,40,60,80\}$;

\item coefficient to be multiplied by the BIC penalty: 


$\lambda \in \{1,1.5,2,2.5,3,4,6,10,20,30\}$;

\item coefficient to be multiplied by the precision matrix in the BGe score:

$a_m \in \{0.0001,0.001,0.01,0.025,0.05,0.1,0.25,0.5,1,1.5\}$, and the degree of freedom $a_w$ in the Wishart distribution is set to be $n + a_m + 1$.

\end{itemize}
Unlike comparing skeletons, the ROC curves created this way can be non-monotonic. Some edges in the patterns may jump between being directed and undirected when the penalty is too small or too large, so the number of TPs may not be monotonically increasing. Furthermore, one can read the SHDs scaled by P directly from the plots, which are the Manhattan distances from the curves to the point $(0,1)$. 

\subsubsection{Simulation Results}
\label{sec:sim}

We generate random DAGs with $4$ expected neighbours per node. The edge weights are uniformly sampled from the intervals $(-1, -0.4) \cup (0.4, 1)$. We consider $3$ network sizes $n \in \{12,20,30\}$ and $3$ sample sizes $N \in \{300,500,800\}$. For each DAG, we generate the associated Gaussian data and perform element-wise discretization randomly using the symmetric Dirichlet distribution $\dir(L_i, \nu)$, where $L_i$ is the number of ordinal levels in the $i^{th}$ dimension and $\nu$ is a concentration parameter. More specifically, we first generate the cell probabilities in the ordinal contingency table of $X_i$ from $\dir(L_i, \nu)$. According to the cell probabilities, we can compute reversely the thresholds for cutting the Gaussian variable $Y_i$ using the normal quantile function (\fref{fig:dirichlet}). Here, we draw $L_i$ randomly from the interval $[2,4]$ to mimic the typical number of ordinal levels and choose $\nu$ to be $2$ to avoid highly skewed contingency tables.  

\begin{figure}[t]
\includegraphics[width = \textwidth]{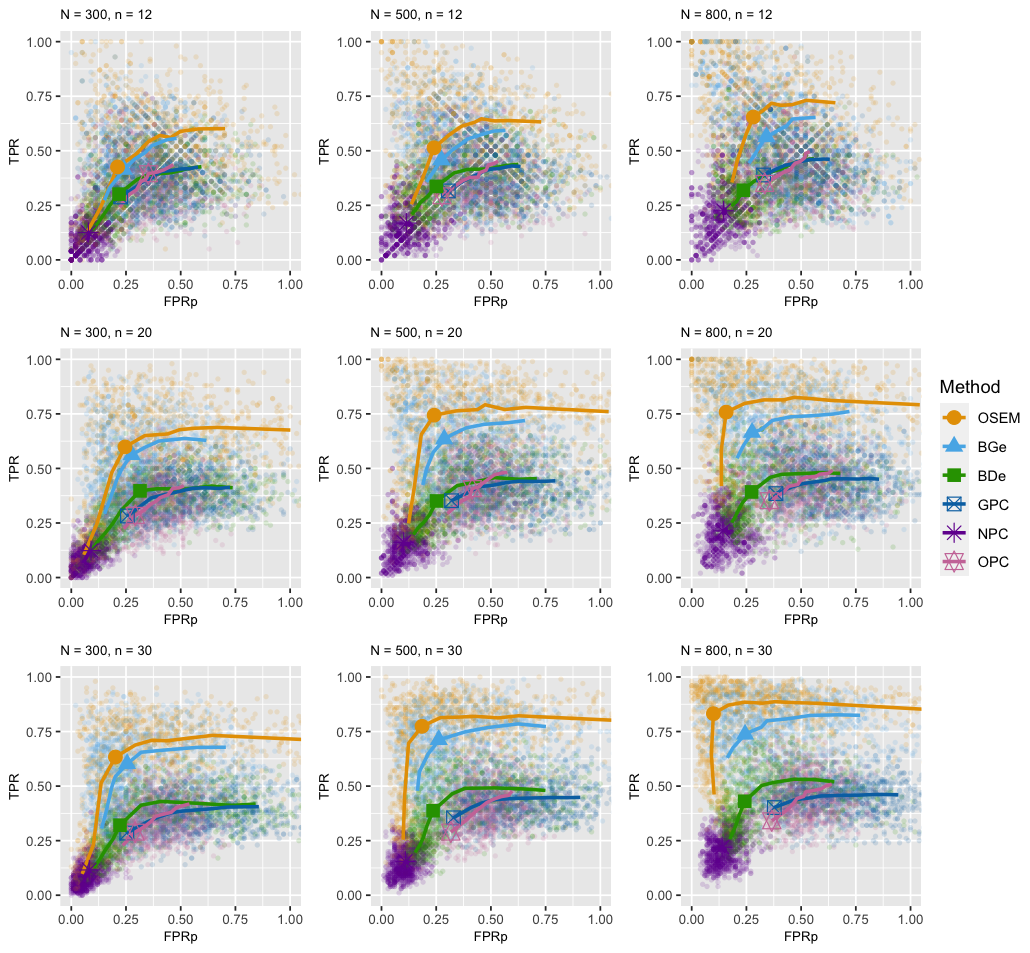}
\caption[Performance comparison in recovering the pattern of the true DAG]{The comparison of the performance in recovering the true pattern between the OSEM algorithm and five other existing approaches: BGe, BDe, GPC, NPC, and OPC. The ROC curves are created by plotting the TPR against the FPRp for sample sizes $N \in \{300,500,800\}$, network sizes $n \in \{12,20,30\}$, $3$ expected number of levels, and a Dirichlet concentration parameter of $2$. Both the x-axis and the y-axis are limited to $[0,1]$. For each method, the point on the line corresponds to the lowest SHD (the highest sum of TPR and $(1 - \text{FPRp})$).}
\label{fig:size_test}
\end{figure}

For each configuration, we repeat $100$ times the process of randomly generating the data and estimating the structure using the six approaches, followed by plotting the ROC curves accordingly. Each point in \fref{fig:size_test} represents a tuple (TPR, FPRp). The lines are created by interpolating the average TPR against the average FPRp at each penalization value. Unlike FPRn, FPRp can be greater than $1$, but only points within $[0,1]$ are shown.

In general, the three PC-based methods alone are too conservative to recover the patterns of the true underlying structures, regardless of how we increase the significance level. Amongst them, the GPC algorithm performs the best, followed by the OPC and NPC algorithms.

While the performance of the hybrid method with the BDe score is sometimes much better than the PC algorithms, the upper bound of the BDe band barely touches the lower bound of the BGe band. This observation is unanticipated in two ways. On one hand, ignoring the ordering amongst the categories is not as harmless as one may expect. If our modeling assumptions are correct, the common practice of using the multinomial distribution cannot be taken for granted, as the estimated DAG can be far off. On the other hand, the BGe score is more powerful than we may have anticipated. Even though treating the ordinal data as continuous is inaccurate, this is still capable of detecting many true conditional independence relationships in the latent space. Thus, leaving aside our OSEM algorithm, one should rather use the BGe score instead of the popular BDe score for Bayesian network learning from ordinal data.

In all cases, our OSEM algorithm demonstrates a strong improvement. When the sample size is large enough, the contingency tables are more accurate, so recovering the original covariance structure through the EM iterations becomes more likely. With fixed sample size, increasing the network size does not make the performance of our model deteriorate, as long as the sparsity of the network stays the same. Additional simulation results can be found in \aref{sec:more_sim}, including 1) ROC curves for skeletons, 2) the effect of thresholds, 3) the effect of network sparsity, 4) runtimes, and 5) comparison against the most recent structure learning methods for mixed data.

\subsection{Predictive Performance}
\label{sec:pred}

For score-based approaches, we also evaluate the predictive performance using five real ordinal data sets from \cite{mcnally2017co} and the UCI machine learning repository \citep{Dua:2019} (\tref{tab:real_data sets}). 

\begin{table}[H]
\small
    \centering
    \begin{tabular}{c|c|c|c}
      \hline
      \textbf{data set} & \textbf{Sample size} & \textbf{Number of variables} & \textbf{Average levels} \\
      \hline
      OCD and Depression & 408 & 24 & 4.67 \\
      Congressional Voting Records & 435 & 17 & 2\\
      Contraceptive Method Choice & 1473 & 9 & 3.3 \\
      Primary Tumor & 339 & 17 & 2.18\\
      SPECT Heart & 267 & 23 & 2\\
      \hline
    \end{tabular}
    \caption{Description of the five real data sets in \sref{sec:pred}. Except the OCD and Depression data set, all other data sets are obtained from the UCI machine learning repository.}
    \label{tab:real_data sets}
\end{table}

\begin{figure}[H]
\centering
\includegraphics[width = \textwidth]{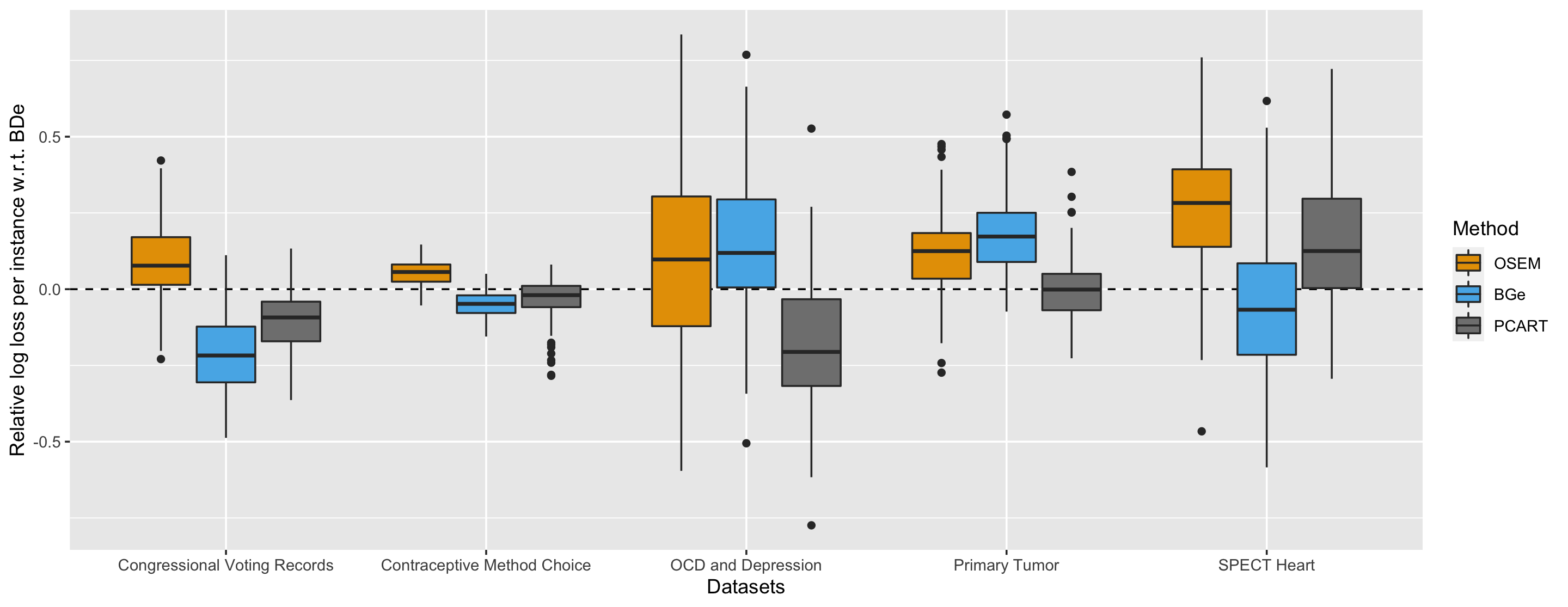}
\caption{The relative log loss per instance with respect to the BDe baseline (dashed horizontal line at $0$) for OSEM, BGe, and PCART over the five data sets and $100$ random splits.}
\label{fig:log_loss}
\end{figure}

In addition to OSEM, BGe, and BDe, we have included the PCART algorithm of \cite{talvitie2019learning} for comparison, which is a score-based method for mixed continuous, nominal and ordinal data. For each data set and method, we train the network with $80\%$ of the data points, and conditioned on the structure, we compute the log loss on the remaining $20\%$ test cases. To ensure that the method with the continuous BGe score has a comparable log loss, we substitute its maximum a posteriori covariance matrix based on the estimated DAG into the OSEM log loss function (\aref{sec:pred_supp}). Taking the BDe as the baseline, which is a standard choice for categorical data, we plot in \fref{fig:log_loss} the relative log loss per instance corresponding to the optimal tuning parameters over $100$ random splits. We can clearly see that OSEM consistently outperforms the baseline approach and is highly competitive across the board, while BGe and PCART are much less robust. This further suggests that modelling the ordinal data as being generated from a hidden continuous space can be more effective in practice, making the OSEM algorithm a generally better choice for both structure learning and prediction. Details including the computation of the log loss, additional figures and more extensive comparisons on sub-sampled data sets can be found in \aref{sec:pred_supp}.

\section{Application to Psychological Data}
\label{sec:psych}

In this section, we provide an example applying our method to psychological survey data, where the potential of Bayesian network models to describe complex relationships has recently gained momentum \citep{mcnally2017co, moffa2017using, kuipers2019links, bird2019adolescent}. We use a data set of size $408$ from a study of the functional relationships between the symptoms of obsessive-compulsive disorder (OCD) and depression \citep{mcnally2017co}. It consists of $10$ five-level ordinal variables representing the OCD symptoms and $14$ four-level ordinal variables representing the depression symptoms, measured with the Yale-Brown Obsessive-Compulsive Scale (Y-BOCS-SR) \citep{steketee1996yale} and the Quick Inventory of Depressive Symptomatology (QIDS-SR) \citep{rush200316} respectively. Details are summarized in \stref{tab:OCDRogers}. Note that ``decreased \vs increased appetite'' and ``weight loss \vs gain'' in the original questionnaire QIDS-SR de facto address the same questions from opposite directions. To avoid including two nodes in the network representing almost identical information, we replace each pair of variables with a one seven--level variable---\textit{appetite} and \textit{weight} respectively. Contradictory answers are merged into the closest levels. For example, if ``decreased appetite'' receives a score of 2 and ``increased appetite'' receives a score of 1, we assigns $-1$ to the variable \textit{appetite}.

\begin{figure}[t]
\centering
\includegraphics[width = \textwidth]{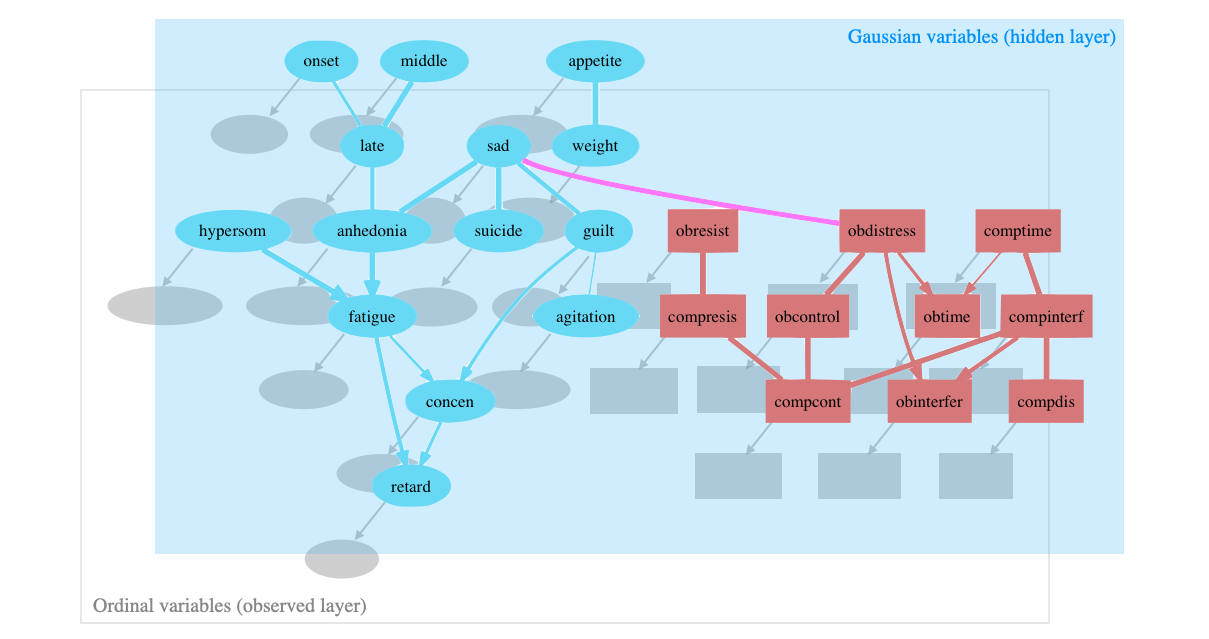}
\caption{CPDAG of the hidden network structure estimated via the OSEM algorithm for the obsessive-compulsive disorder (OCD) symptoms (rectangles) and depression symptoms (ellipses) from a data set provided by \cite{mcnally2017co}. The labelled nodes are the latent Gaussian variables, and the shaded nodes are the corresponding observed ordinal variables. The Monte Carlo sample size $K$ is $5$, and the penalty coefficient $\lambda$ is $6$. The thickness of the undirected edges reflects the percentage of time they occur in the skeletons of $500$ Bootstrapped CPDAGs. If an edge between two labelled nodes is directed, then its thickness represents its directional strength. For clarity, we highlight the bridge between the two clusters of nodes (\textit{sad} -- \textit{obdistress}).}
\label{fig:psych_OSEM}
\end{figure}

In addition to having a suitable number of observations $N = 408$, variables $n = 24$, and an average number of levels greater than $4$, we also check that none of the marginal contingency tables have weights concentrating on one end. Thus, our OSEM algorithm is particularly well suited in this setting. We choose the Monte Carlo sample size $K$ to be $5$ and the penalty coefficient $\lambda$ to be $6$, which corresponds to the highest sum of average TPR and $(1 - \text{FPRp})$ on the ROC curves for $N = 500$, $n = 20$ or $30$, $\mathbb{E}[L_i] = 4$ or $5$, and $\nu = 1$ from our simulations studies. The resulting CPDAG is shown in \fref{fig:psych_OSEM}. We depict the nodes related to OCD symptoms with rectangles and the nodes related to depression symptoms with ellipses. The thickness of the undirected edges reflects the percentage of time they occur in the skeletons of $500$ bootstrapped CPDAGs. If an edge is directed, then its thickness represents its directional strength. 

We reproduce the Bayesian network of \cite{mcnally2017co} by running the hill-climbing algorithm with the Gaussian BIC score and retaining the edges that appear in at least $85\%$ of the $500$ Bootstrapped networks. The corresponding CPDAG is depicted in \sfref{fig:psych_mcnally}. In both graphs, the OCD and depression symptoms form separate clusters, bridged by the symptoms of sadness and distress caused by obsessions. With the exception of a few edge discrepancies (\eg \textit{late} -- \textit{anhedonia}, \textit{guilt} $\rightarrow$ \textit{suicide}), the two graphs appear to be mostly similar, which is consistent with our simulation results. When the average number of ordinal levels is relatively large, violating the hidden Gaussian assumption does not substantially reduce the performance in recovering the true edges. Nevertheless, the structure estimated by the OSEM algorithm should still be more reliable. 

Moreover, the CPDAG estimated using the hybrid method with the BDe score (\sfref{fig:psych_BDe}) is also very similar to the OSEM output, with some edges missing and all edges undirected. The drastic difference in the estimated structures between the BDe and OSEM observed in the simulation studies does not recur here. A potential reason could be that the average number of parents in the true network is fewer than one. In this case, one can check via simulations that the BDe score can sometimes provide satisfactory results without the problem of overparameterization. However, we still observe a general improvement in predictive power for OSEM over the BDe approach (\fref{fig:log_loss}). Also, the OSEM algorithm can handle networks that are less sparse and hence should be more robust and preferable in practical applications. \sfref{fig:psych_BGe} is the CPDAG estimated using the hybrid method with the BGe score. Again, since it treats the data as continuous and uses Gaussian distribution, it looks similar to the first two networks but slightly denser. Filtering out edges that are less certain may improve the learned structure, but the outcome is unlikely to outperform the one from the OSEM algorithm. 

Finally, we use heatmaps (\sfref{fig:heatmap}) to visualize and compare the adjacency matrices of the four CPDAGs. The shade in the grid represents the percentage of time a directed edge appears in the $500$ Bootstrapped CPDAGs, where an undirected edge counts half for each direction. The darker the shade, the more frequent the corresponding directed edge occurs. There are several aspects in common: first, the symptoms of the two questionnaires form relatively separate clusters in all four networks; second, the symptoms \textit{appetite} and \textit{weight} interact with each other but are isolated from the rest of the symptoms; third, the connection between \textit{sad} and \textit{obdistress} is present in all four structures. Nevertheless, there are also small differences between the network structures which may be important in practice. For example, the link \textit{late} -- \textit{anhedonia} creates a connection between a small subset of symptoms and the largest cluster, a connection which we do not observe in the network estimated by \cite{mcnally2017co}. Since our simulations suggest that the OSEM achieves better performances in structure learning from ordinal data, discovering the connections more accurately may be highly relevant in the application domain.

\section{Discussion}
\label{sec:discussion}

In this work, we addressed the problem of learning Bayesian networks from ordinal data by combining the multinomial probit model \citep{daganzo2014multinomial} with the Structural EM algorithm of \cite{friedman1997learning}. The resulting framework is the Ordinal Structural EM (OSEM) algorithm. By assuming that each ordinal variable is obtained by marginally discretizing an underlying Gaussian variable, we can capture the ordinality amongst the categories. By contrast, the commonly used multinomial distribution loses information related to the order of categories due to its invariance with respect to any random permutation of the levels. Furthermore, we assumed that the hidden variables jointly follow a DAG structure. In other words, we have a Gaussian DAG model in the latent space, allowing us to exploit many of its well-established properties, such as decomposability and score-equivalence. 

The location and scale of the original Gaussian distribution are no longer identifiable after discretization. Hence, we chose to standardize each latent dimension to ensure identifiability. This is computationally efficient since we only need to estimate the thresholds once in the process. 

Instead of a random initialization, we started with a full DAG and used a pairwise likelihood approach for estimating the initial correlation matrix. This method reduces the number of EM iterations required for convergence. We also derived closed-form formulas for both the structure and parameter updates. In the parameter update, we can perform regression with subset selection following the topological order of a DAG, which induces little cost. Since our expected scoring function is decomposable, score-equivalent, and consistent, we can perform the structure update efficiently by using existing search schemes, such as the GES, MCMC samplers, and hybrid approaches. Unlike the reversible jump MCMC method of \cite{webb2008bayesian}, this additional flexibility in choosing search methods allows us to further reduce the computational burden. 

In the simulation studies, we compared the OSEM algorithm to other existing approaches, including three PC-based methods, two hybrid methods with the BDe and the BGe scores respectively, and several methods for mixed data. Under all configurations we tested, our method significantly improved the accuracy in recovering the hidden DAG structure. Using real ordinal data sets, we also showed the generally superior predictive performance of the OSEM algorithm over other score-based approaches, highlighting the usefulness of our modelling assumptions in real applications. In addition, we demonstrated the practicality of our method by applying it to psychological survey data.

To our surprise, the performance of the BDe score in structure recovery was much poorer than our method. Even though it is widely recognized that ignoring the ordering can lead to loss of information, we are, to the best of our knowledge, the first to quantify and visualize its impact on structural inference. The resulting deviance from the true structure was much larger than expected. Therefore, we should in general avoid using nominal methods to tackle problems involving ordinal data. 

The BGe score, on the other hand, performed better than we had thought. The correlation matrix estimated directly from the ordinal data appears to be sufficient for the BGe score to identify many of the true edges, especially when the number of ordinal levels is large, so that the continuous approximation improves. However, this method can hardly go beyond the OSEM algorithm, since it still disregards the discreteness of the data.

The most dominating factor for successful structure learning is the quality of the contingency tables, which is determined by the sample size, the number of levels in each ordinal variable, as well as the position of the thresholds. First, the sample size is tightly associated with the accuracy of each cell probability, which decides to what extent the original correlation matrix can be recovered. Second, the number and position of the thresholds control the resolution of the contingency tables. In order to obtain a more reliable result, it is recommended to have more than two levels for each dimension, and the weight should not concentrate on one end of the tables. Otherwise, one may reconsider the effectiveness of the data collected, such as whether a survey question is well-designed or whether the target population selected is appropriate.

The PC algorithms have relatively limited accuracy in estimating the latent structures. It could be interesting to see if a more suitable test for ordinal data can be developed, which is in itself a challenging task. A recent method proposed by \cite{liu2020assessing} may be a direction to investigate further. Furthermore, a more accurate output by the PC algorithm can also refine the initial search space for a hybrid search scheme, such as the method of \cite{kuipers2018efficient}, which should in principle bring further improvements to our OSEM algorithm.

Because of the EM iterations, the runtime of our method is unavoidably longer than a pure score-based or hybrid approach, especially when the network size gets larger. The main bottleneck lies in the sampling from the truncated multivariate normal distribution, which is required in every iteration of the process. A possible direction for improvement is a Sequential Monte Carlo sampler \citep{moffa2014sequential}, where the samples can be updated from one iteration to the next instead of resampling from scratch. 

Another possible direction is to create a Bayesian version of the model. One can replace the BIC penalty with the well-known Wishart prior \citep{geiger2002parameter} and adapt the Bayesian Structural EM algorithm of \cite{friedman1998structural}. Moreover, Bayesian model averaging \citep{madigan1994model, friedman2003being} may help with the problem of reaching a local optimum. Specifically, at each iteration, one can sample a collection of structures instead of checking only the highest-scoring graph. A sequential Monte Carlo method can also be applied to the structure level to overcome the computational bottleneck.

Under our current model specification, it is natural to relate to the problem of Bayesian network learning with mixed data. In particular, one may obtain a data set with both continuous and ordinal variables by first generating a Gaussian data set according to a DAG structure and then discretizing some of the variables while keeping others continuous. A similar learning framework may be applicable in this setting. The situation, however, can become much more sophisticated when one wants to include nominal categorical variables. It would therefore be interesting to extend OSEM to mixed continuous and categorical data and compare with other existing approaches \citep{cui2016copula, tsagris2018constraint, talvitie2019learning}.


\acks{The authors would like to thank Niko Beerenwinkel for helpful comments and discussions and for hosting Xiang Ge Luo during her Master's Thesis. We also wish to thank the two anonymous reviewers and the editor for their careful reviews and constructive feedback.}


\appendix

\setcounter{section}{0} 

\renewcommand\thefigure{S\arabic{figure}}
\setcounter{figure}{0} 
\renewcommand\thetable{S\arabic{table}}
\setcounter{table}{0} 

\section{The OSEM Algorithm}
\label{sec:algorithm}

\begin{algorithm}[H]
\label{algo:OSEM}
\SetKwRepeat{Repeat}{loop}{until}
\SetKwFor{Init}{initialization:}{}{}
\SetKwFor{For}{for}{}{}
\SetAlgoLined
 \Init{}{
 	Let $\mathcal{G}^{(0)}$ be a full graph\;
 	Estimate thresholds $\hat{\bm{\alpha}}$: $\hat{\alpha}(i,l) = \Phi^{-1} \Big( \frac{1}{N} \sum^N_{j = 1} \mathds{1}(x^j_i \leq \tau(i,l)) \Big)$, $l = 1, \dots, L_i -1 $\;
	Estimate correlation matrix $\Sigma^{(0)}$ such that for all $i,j \in \{1,\dots,n\}$, $i \neq j,$
	\[
\rho^{(0)}_{ij} \mid \hat{\bm{\alpha}} = \text{argmax}_{\rho_{ij} \in (-1,1)}  \log p \bigg(\mathcal{D}_{X_i \cup X_j} \mid 
\begin{bmatrix}
0 \\
0
\end{bmatrix}
,
\begin{bmatrix}
1 &  \rho_{ij}\\
\rho_{ij} & 1
\end{bmatrix}
, \mathcal{G}^{(0)} \bigg);
	\]
 	
 }
 \Repeat($: t = 0,1,2,\dots$){convergence}{
	Sample $K$ hidden data points for each observation $\mathbf{x}^j$ from the distribution
	\[
	\mathbf{Y}^j \mid \mathbf{x}^j, \hat{\bm{\alpha}}, \Sigma^{(t)}, \mathcal{G}^{(t)} \sim \TMN(\bm{0},\Sigma^{(t)} \mid \mathbf{x}^j, \hat{\bm{\alpha}});
	\] 	
 	\underline{\textbf{Structure update}}:
	
	Compute expected covariance matrix 
	\[	
	\hat{\Sigma} = \frac{1}{N}\frac{1}{K} \sum^N_{j = 1} \sum^K_{k = 1} (\mathbf{y}^{j(k)}) (\mathbf{y}^{j(k)})^{\top};
	\]
	
 	Update the DAG structure
	\[
	\begin{split}
	\mathcal{G}^{(t+1)} & = \argmax_{\mathcal{G}: \text{ DAG over } (\mathbf{X},\mathbf{Y})} \sum^n_{i = 1} \bigg( -\frac{N}{2} \bigg[\log \Big(\hat{\Sigma}_{i,i} - \hat{\Sigma}_{i,pa(i)} \hat{\Sigma}_{pa(i),pa(i)}^{-1} \hat{\Sigma}_{pa(i),i} \Big) \bigg] \\
	& \qquad \qquad \qquad \qquad \qquad \; \; - \frac{\log N}{2} \dim(Y_i,\mathbf{Y}_{pa(i)}) \bigg);
	\end{split}
	\]

  	\underline{\textbf{Parameter update}}:
	
	Following the topological order in $\mathcal{G}^{(t+1)}$, update the parameters using linear regression with best subset selection and the BIC penalty:
	\[
	\begin{split}
	& \mathbf{b}^{(t+1)}_i = \bigg(\sum^K_{k = 1} \Big(\mathbf{Y}_{pa(i)}^{(k)} \Big)^{\top} \mathbf{Y}_{pa(i)}^{(k)} \bigg)^{-1} \bigg(\sum^K_{k = 1} \Big(\mathbf{Y}_{pa(i)}^{(k)} \Big)^{\top} \mathbf{y}^{(k)}_i \bigg); \\
	& v^{(t+1)}_i = \frac{1}{N} \frac{1}{K} \sum_{j = 1}^N  \sum^K_{k = 1} \Big( y^{j(k)}_i - (\mathbf{b}^{(t+1)}_i)^{\top} \mathbf{y}^{j(k)}_{pa(i)} \Big)^2;
	\end{split}
	\]

	Transform $\{\mathbf{b}^{(t+1)}_i, v^{(t+1)}_i\}^n_{i = 1}$ into the correlation matrix $\Sigma^{(t+1)}$\;
 	
 }
 \caption{Ordinal Structural EM (OSEM)}
\end{algorithm}

\subsection{Threshold Estimation}

We only need to estimate the thresholds once at the beginning of the EM iterations, because the discretization is performed marginally, and the marginal mean and variance are always fixed. Consider the $i^{th}$ dimension. By the law of large numbers and the law of iterated expectation, for all $l = 1,\dots, L_i$,
\[
\begin{split}
\frac{1}{N} \sum^N_{j = 1} \mathds{1}(X^j_i = \tau(i,l)) & \rightarrow \mathbb{E} \big[\mathds{1}(X_i = \tau(i,l)) \big] \\
& = \mathbb{E} \big[ \mathbb{E} \big[\mathds{1}(X_i = \tau(i,l)) \mid Y_i \big] \big] \\
& = \mathbb{E} \big[ P(X_i = \tau(i,l) \mid Y_i, \bm{\alpha}_i) \big] \\
& = \mathbb{E} \big[ \mathds{1}( Y_i \in [\alpha(i,l-1), \alpha(i,l)) ) \big] \\
& = P \big( Y_i \in [\alpha(i,l-1), \alpha(i,l)) \big),
\end{split}
\]
as $N \rightarrow \infty$. It follows that
\[
\frac{1}{N} \sum^N_{j = 1} \mathds{1}(X^j_i \leq \tau(i,l)) \rightarrow P(Y_i < \alpha(i,l)) = \Phi \big(\alpha(i,l) \big), \qquad N \rightarrow \infty.
\]
Thus, given a sample $\mathcal{D}_{\mathbf{X}} = \{\mathbf{x}^1, \dots, \mathbf{x}^N\}$, we can estimate each threshold $\alpha(i,l)$ as
\[
\hat{\alpha}(i,l) = \Phi^{-1} \Big( \frac{1}{N} \sum^N_{j = 1} \mathds{1}(x^j_i \leq \tau(i,l)) \Big), \qquad l = 1,\dots, L_i - 1.
\]
They are simply the empirical quantiles of the standard normal distribution.

\subsection{Score Used in the Structure Update}

Let $P(i) := pa(i)$ and $O(i) := pa(i) \cup \{i\}$ for all $i \in \{1,\dots,n\}$. From \eref{eq:model_ordinal}, the joint probability of $\mathbf{x}$ and $\mathbf{y}$ can be written as
\[
\begin{split}
p(\mathbf{x},\mathbf{y} \mid \theta, \mathcal{G}) & = \prod^n_{i = 1} \phi(y_i \mid \mathbf{y}_{P(i)}, \vartheta_i, \mathcal{G})p(x_i \mid y_i, \bm{\alpha}_i) \\
& = \prod^n_{i = 1} \frac{\phi(\mathbf{y}_{O(i)} \mid (\bm{\mu},\Sigma)_{O(i)}) }{\phi(\mathbf{y}_{P(i)} \mid (\bm{\mu},\Sigma)_{P(i)}) } p(x_{i} \mid y_{i} , \bm{\alpha}_i),
\end{split}
\]
where $(\bm{\mu},\Sigma)_{O(i)}$ is the set of parameters indexed by $O(i)$. The $Q$ function becomes
\[
Q(\mathcal{G}, \theta; \mathcal{G}^{(t)}, \theta^{(t)}) = \sum_{j = 1}^N \mathbb{E}_{\mathbf{Y}^j \mid \mathbf{x}^j, \mathcal{G}^{(t)}, \theta^{(t)}}\Big[\log p(\mathbf{X}^j, \mathbf{Y}^j \mid \mathcal{G}, \theta) \mid \mathbf{X}^j = \mathbf{x}^j \Big]
\]
\[
\begin{split}
& = \sum_{j = 1}^N \sum^n_{i = 1} \mathbb{E}_{\mathbf{Y}^j \mid \mathbf{x}^j,\mathcal{G}^{(t)},  \theta^{(t)}} \Big[\log \frac{\phi(\mathbf{Y}^j_{O(i)} \mid (\bm{\mu},\Sigma)_{O(i)}) }{\phi(\mathbf{Y}^j_{P(i)} \mid (\bm{\mu},\Sigma)_{P(i)}) } p(X^j_{i} \mid Y^j_{i}, \bm{\alpha}_i) \mid X^j_i = x^j_i , \mathcal{G}\Big] \\
& = \sum^n_{i = 1} \sum_{j = 1}^N \mathbb{E}_{\mathbf{Y}^j \mid \mathbf{x}^j,\mathcal{G}^{(t)},  \theta^{(t)}} \Big[\log \phi(\mathbf{Y}^j_{O(i)} \mid (\bm{\mu},\Sigma)_{O(i)}) p(X^j_{i} \mid Y^j_{i}, \bm{\alpha}_i) \mid X^j_i = x^j_i , \mathcal{G}\Big] \\
& \qquad \qquad \  - \mathbb{E}_{\mathbf{Y}^j \mid \mathbf{x}^j,\mathcal{G}^{(t)},  \theta^{(t)}} \Big[\log \phi(\mathbf{Y}^j_{P(i)} \mid (\bm{\mu},\Sigma)_{P(i)}) \mid \mathcal{G}\Big].
\end{split}
\]
Using the multivariate normal density $\phi$ and the truncated multivariate normal density $\varphi$, we can expand the summands,
\[
\begin{split}
& \qquad \mathbb{E}_{\mathbf{Y}^j \mid \mathbf{x}^j,\mathcal{G}^{(t)},  \theta^{(t)}} \Big[\log \phi(\mathbf{Y}^j_{O(i)} \mid (\bm{\mu},\Sigma)_{O(i)}) p(X^j_{i} \mid Y^j_{i}, \bm{\alpha}_i) \mid X^j_i = x^j_i , \mathcal{G} \Big] \\
& = \int \Big(\log \phi(\mathbf{y}^j_{O(i)} \mid (\bm{\mu},\Sigma)_{O(i)}) \Big) \varphi(\mathbf{y}^j \mid \theta^{(t)},\mathcal{G}^{(t)}, \bm{\alpha}) d \mathbf{y}^j \\
& = -\frac{1}{2} \int \Big[ (\lvert P(i) \rvert + 1) \log(2\pi) + \log(\lvert \Sigma_{O(i),O(i)} \rvert) \\
& \qquad \qquad + \Tr \big((\mathbf{y}^j_{O(i)} - \bm{\mu}_{O(i)}) (\mathbf{y}^j_{O(i)} - \bm{\mu}_{O(i)} )^{\top} \Sigma_{O(i),O(i)}^{-1} \big) \Big] \varphi(\mathbf{y}^j \mid \theta^{(t)},\mathcal{G}^{(t)}, \bm{\alpha}) d \mathbf{y}^j \\
& = -\frac{1}{2} \Big[(\lvert P(i) \rvert + 1) \log(2\pi) + \log(\lvert \Sigma_{O(i),O(i)} \rvert) \\
& \qquad \qquad + \Tr \big(\mathbb{E}_{\mathbf{Y}^j \mid \mathbf{x}^j,\mathcal{G}^{(t)},  \theta^{(t)}}[(\mathbf{Y}^j_{O(i)} - \bm{\mu}_{O(i)}) (\mathbf{Y}^j_{O(i)} - \bm{\mu}_{O(i)} )^{\top}] \Sigma_{O(i),O(i)}^{-1}\big) \Big].
\end{split} 
\]
Analogously,
\[
\begin{split}
& \mathbb{E}_{\mathbf{Y}^j \mid \mathbf{x}^j,\mathcal{G}^{(t)},  \theta^{(t)}} \Big[\log \phi(\mathbf{Y}^j_{P(i)} \mid (\bm{\mu},\Sigma)_{P(i)}) \mid \mathcal{G}\Big] = -\frac{1}{2} \Big[\lvert P(i) \rvert \log(2\pi) + \log(\lvert \Sigma_{P(i),P(i)} \rvert) \\
& \qquad \qquad + \Tr \big(\mathbb{E}_{\mathbf{Y}^j \mid \mathbf{x}^j,\mathcal{G}^{(t)},  \theta^{(t)}}[(\mathbf{Y}^j_{P(i)} - \bm{\mu}_{P(i)}) (\mathbf{Y}^j_{P(i)} - \bm{\mu}_{P(i)} )^{\top}] \Sigma_{P(i),P(i)}^{-1}\big) \Big].
\end{split}
\]
By computing the expected statistics
\[
\hat{\bm{\mu}} = \frac{1}{N} \sum^N_{j = 1} \mathbb{E}_{\mathbf{Y}^j \mid \mathbf{x}^j,\mathcal{G}^{(t)},  \theta^{(t)}} \big[\mathbf{Y}^j \big]
\]
and
\[
\hat{\Sigma} = \frac{1}{N} \sum^N_{j = 1} \mathbb{E}_{\mathbf{Y}^j \mid \mathbf{x}^j,\mathcal{G}^{(t)},  \theta^{(t)}} \big[(\mathbf{Y}^j - \hat{\bm{\mu}}) (\mathbf{Y}^j - \hat{\bm{\mu}})^{\top} \big],
\]
we get
\[
Q(\mathcal{G}, \hat{\theta} ; \mathcal{G}^{(t)}, \theta^{(t)})  = \sum^n_{i = 1} -\frac{N}{2} \bigg[\log(2\pi) + \log \Big(\frac{\lvert \hat{\Sigma}_{O(i),O(i)} \rvert}{\lvert \hat{\Sigma}_{P(i),P(i)} \rvert} \Big) + 1 \bigg].
\]
Notice that $\lvert \hat{\Sigma}_{O(i),O(i)} \rvert$ can be partitioned as
\[
\hat{\Sigma}_{O(i),O(i)}
= 
\begin{bmatrix}
\hat{\Sigma}_{i,i} & \hat{\Sigma}_{i,P(i)} \\
\hat{\Sigma}_{P(i),i} & \hat{\Sigma}_{P(i),P(i)}
\end{bmatrix}
\]
with 
\[
\lvert \hat{\Sigma}_{O(i),O(i)} \rvert = \lvert \hat{\Sigma}_{P(i),P(i)} \rvert (\hat{\Sigma}_{i,i} - \hat{\Sigma}_{i,P(i)} \hat{\Sigma}_{P(i),P(i)}^{-1} \hat{\Sigma}_{P(i),i}).
\]
Therefore,
\[
Q(\mathcal{G}, \hat{\theta} ; \mathcal{G}^{(t)}, \theta^{(t)}) = \sum^n_{i = 1} -\frac{N}{2} \bigg[\log \Big(\hat{\Sigma}_{i,i} - \hat{\Sigma}_{i,P(i)} \hat{\Sigma}_{P(i),P(i)}^{-1} \hat{\Sigma}_{P(i),i} \Big) + \log(2\pi) + 1 \bigg].
\]
The formulas used in \alref{algo:OSEM} are after the adjustment for the identifiability constraints.

\subsection{MLE in the Parameter Update}

By substituting \eref{eq:model_ordinal} into $Q(\mathcal{G}, \theta; \mathcal{G}^{(t)}, \theta^{(t)})$, we get
\[
\begin{split}
Q(\mathcal{G}, \theta; \mathcal{G}^{(t)}, \theta^{(t)}) & = \sum_{j = 1}^N \mathbb{E}_{\mathbf{Y}^j \mid \mathbf{x}^j, \mathcal{G}^{(t)}, \theta^{(t)}}\Big[\log p(\mathbf{X}^j, \mathbf{Y}^j \mid \mathcal{G}, \theta) \mid \mathbf{X}^j = \mathbf{x}^j \Big] \\
& = \sum_{j = 1}^N \sum^n_{i = 1} \mathbb{E}_{\mathbf{Y}^j \mid \mathbf{x}^j,\mathcal{G}^{(t)},  \theta^{(t)}} \Big[\log \phi(Y^j_i \mid \mathbf{Y}^j_{pa(i)}, \theta_i, \mathcal{G}) p(X^j_i \mid Y^j_i, \bm{\alpha}_i) \mid X^j_i = x^j_i \Big] \\
& = \sum^n_{i = 1} \sum_{j = 1}^N \int \Big(\log \phi(y^j_i \mid \mathbf{y}^j_{pa(i)}, \theta_i, \mathcal{G}) \Big) \varphi(\mathbf{y}^j \mid \theta^{(t)},\mathcal{G}^{(t)}, \bm{\alpha}) d \mathbf{y}^j \\
& = \sum^n_{i = 1} \sum_{j = 1}^N \mathbb{E}_{\mathbf{Y}^j \mid \mathbf{x}^j, \mathcal{G}^{(t)}, \theta^{(t)}}\Big[\log \phi(Y^j_i \mid \mathbf{Y}^j_{pa(i)}, \theta_i, \mathcal{G})  \Big].
\end{split}
\]
Conditioned on $\theta^{(t+1)}_{pa(i)}$, let $\bm{\beta}_i = (\mu_i, \mathbf{b}_i^{\top})^{\top}$ and $\tilde{\mathbf{Y}}^j_{pa(i)} = (1,(\mathbf{Y}^j_{pa(i)} - \bm{\mu}^{(t+1)}_{pa(i)})^{\top})^{\top} $. Invoking the formula for normal distributions,
\[
\begin{split}
Q(\mathcal{G}^{(t+1)}, \theta; \mathcal{G}^{(t)}, \theta^{(t)}) & = \sum^n_{i = 1} \sum_{j = 1}^N \int \bigg[-\frac{1}{2}\log(2\pi v_i) - \frac{1}{2v_i}\Big(y^j_i - \mu_i - \mathbf{b}_i^{\top} (\mathbf{y}^j_{pa(i)} - \bm{\mu}_{pa(i)}) \Big)^2 \bigg] \\
& \qquad \qquad \qquad \times \varphi(\mathbf{y}^j \mid \theta^{(t)},\mathcal{G}^{(t)}, \bm{\alpha}) d \mathbf{y}^j
\end{split}
\]
\[
\begin{split}
& = \sum^n_{i = 1} - \frac{N}{2} \bigg(\log(2\pi v_i)  + \frac{1}{v_i} \frac{1}{N} \sum_{j = 1}^N \mathbb{E}_{\mathbf{Y}^j \mid \mathbf{x}^j,\mathcal{G}^{(t)},  \theta^{(t)}} \Big[\big(Y^j_i - \mu_i - \mathbf{b}_i^{\top} (\mathbf{Y}^j_{pa(i)} - \bm{\mu}_{pa(i)}) \big)^2 \Big] \bigg) \\
& = \sum^n_{i = 1} - \frac{N}{2} \bigg(\log(2\pi v_i)  + \frac{1}{v_i} \frac{1}{N} \sum_{j = 1}^N \mathbb{E}_{\mathbf{Y}^j \mid \mathbf{x}^j,\mathcal{G}^{(t)},  \theta^{(t)}} \Big[\big(Y^j_i - \bm{\beta}_i^{\top} \tilde{\mathbf{Y}}^j_{pa(i)}  \big)^2 \Big]  \bigg) .
\end{split}
\]
The expectations can be approximated with Monte Carlo samples of size K,
\[
\mathbb{E}_{\mathbf{Y}^j \mid \mathbf{x}^j,\mathcal{G}^{(t)},  \theta^{(t)}} \Big[\big(Y^j_i - \bm{\beta}_i^{\top} \tilde{\mathbf{Y}}^j_{pa(i)}  \big)^2 \Big] \approx \frac{1}{K} \sum^K_{k = 1} \Big[\big(y^{j(k)}_i - \bm{\beta}_i^{\top} \tilde{\mathbf{y}}^{j(k)}_{pa(i)}  \big)^2 \Big],
\]
where $\mathbf{y}^{j(k)}$ is drawn from the conditional distributions $\mathbf{Y}^j \mid \mathbf{x}^j,\mathcal{G}^{(t)},  \theta^{(t)}$, $j = 1, \dots, N$.
Next, we differentiate $Q(\mathcal{G}^{(t+1)}, \theta; \mathcal{G}^{(t)}, \theta^{(t)})$ with respect to $\bm{\beta}_i$ and set to zero,
\[
\begin{split}
\frac{\pd}{\pd \bm{\beta}_i} Q(\mathcal{G}^{(t+1)}, \theta; \mathcal{G}^{(t)}, \theta^{(t)}) & = \frac{\pd}{\pd \bm{\beta}_i} \frac{1}{v_i} \frac{1}{N} \frac{1}{K} \sum_{j = 1}^N \sum^K_{k = 1} \Big[\big(y^{j(k)}_i - \bm{\beta}_i^{\top} \tilde{\mathbf{y}}^{j(k)}_{pa(i)}  \big)^2 \Big] \\
& = -2 \frac{1}{v_i} \frac{1}{N} \frac{1}{K} \sum_{j = 1}^N \sum^K_{k = 1} \Big[\big(y^{j(k)}_i - \bm{\beta}_i^{\top} \tilde{\mathbf{y}}^{j(k)}_{pa(i)}  \big) \tilde{\mathbf{y}}^{j(k)}_{pa(i)} \Big] \\
& \triangleq 0.
\end{split}
\]
Thus,
\[
\bm{\beta}^{(t+1)}_i = \bigg(\sum^K_{k = 1} \Big(\tilde{\mathbf{Y}}_{pa(i)}^{(k)} \Big)^{\top} \tilde{\mathbf{Y}}_{pa(i)}^{(k)} \bigg)^{-1} \bigg(\sum^K_{k = 1} \Big(\tilde{\mathbf{Y}}_{pa(i)}^{(k)} \Big)^{\top} \mathbf{y}^{(k)}_i \bigg)
\]
with $\tilde{\mathbf{Y}}_{pa(i)}^{(k)} = (\tilde{\mathbf{y}}^{1(k)}_{pa(i)} ,\dots, \tilde{\mathbf{y}}^{N(k)}_{pa(i)} )^{\top} \in \mathbb{R}^{N \times (\lvert pa(i) \rvert +1 )}$ and $\mathbf{y}^{(k)}_i = (y^{1(k)}_i, \dots, y^{N(k)}_i)^{\top} \in \mathbb{R}^N$.
Conditioned on $\bm{\beta}^{(t+1)}_i$, we then differentiate $Q(\mathcal{G}^{(t+1)}, \theta; \mathcal{G}^{(t)}, \theta^{(t)})$ with respect to $v_i$,
\[
\frac{\pd}{\pd v_i} Q(\mathcal{G}^{(t+1)}, \theta; \mathcal{G}^{(t)}, \theta^{(t)}) = \frac{1}{v_i} - \frac{1}{v_i^2} \frac{1}{N} \frac{1}{K} \sum_{j = 1}^N \sum^K_{k = 1} \Big[\big(y^{j(k)}_i - \bm{\beta}_i^{\top} \tilde{\mathbf{y}}^{j(k)}_{pa(i)}  \big)^2 \Big] \triangleq 0.
\]
Therefore, the update for $v_i$ is
\[
v^{(t+1)}_i = \frac{1}{N} \frac{1}{K} \sum_{j = 1}^N  \sum^K_{k = 1} \Big[\big(y^{j(k)}_i - (\bm{\beta}^{(t+1)}_i)^{\top} \tilde{\mathbf{y}}^{j(k)}_{pa(i)}  \big)^2 \Big].
\]
Again, the formulas used in \alref{algo:OSEM} are adjusted for the identifiability constraints.

\section{Additional Experimental Results}
\label{sec:more_sim}

\subsection{ROC Curves for Skeletons}

\begin{figure}[t]
\centering
\includegraphics[width = \textwidth]{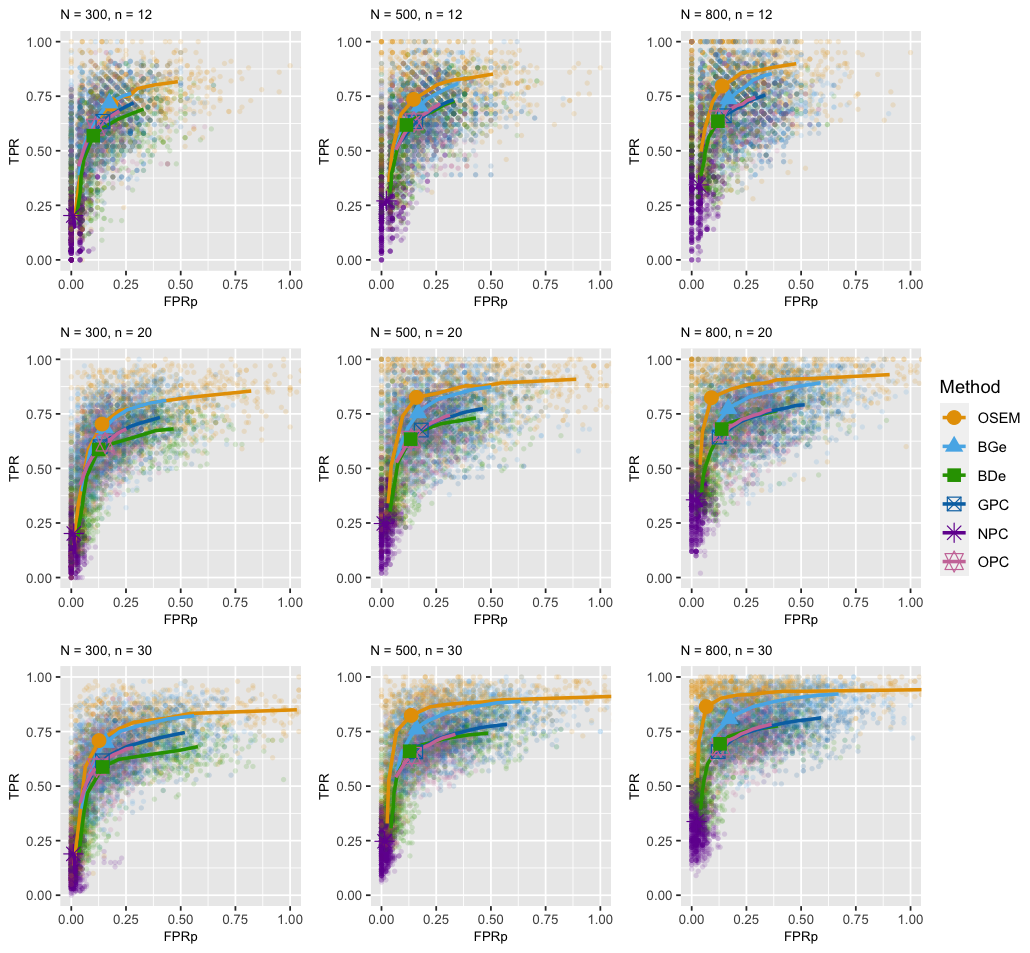}
\caption{The same plot as Figure~\ref{fig:size_test} but the TPR and FPRp are computed using the skeletons instead of the patterns.}
\label{fig:sizes_skeleton}
\end{figure}

If we plot the ROC curves for skeletons, ignoring the v-structure differences, we can see in \fref{fig:sizes_skeleton} that all methods show substantial improvement, and the OSEM algorithm still performs the best. In particular, the method with the BDe score, the GPC and the OPC algorithms have more significant increase in TPR from the one in \fref{fig:size_test}, which implies that these methods tend to inaccurately estimate the v-structures.

\subsection{Effect of the Thresholds}
\label{sec:thresholds}

Here we focus on the effects of thresholds on the model performance by changing the expected number of levels and the symmetric Dirichlet concentration parameter. We pick networks with $20$ nodes, $4$ expected neighbours, and the associated data sets of size $500$.

\begin{figure}[H]
\centering
\includegraphics[width = \textwidth]{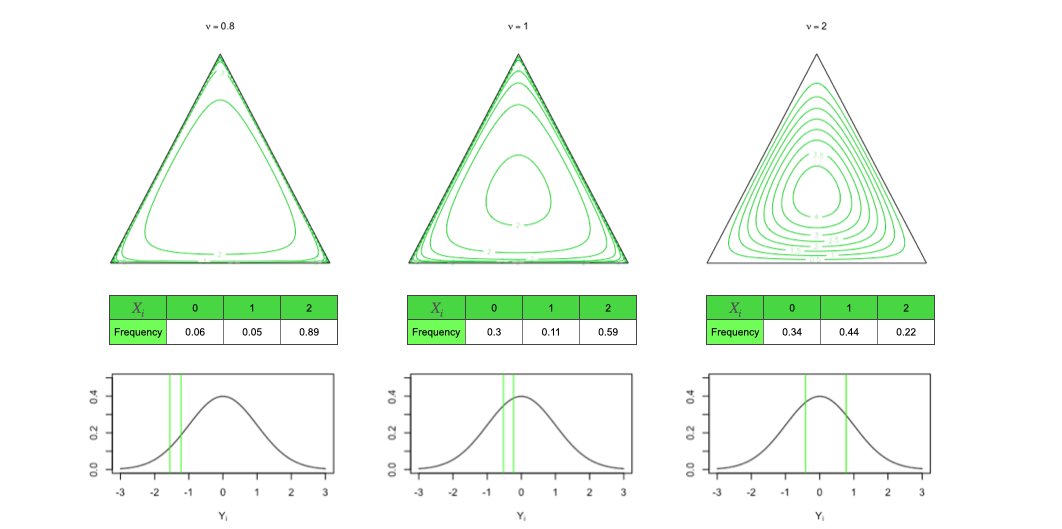}
\caption[Dirichlet distributions]{Contour plots of 3-dimensional symmetric Dirichlet distributions $\dir(3, \nu)$ with $\nu$ being $0.8$ (left), $1 + 1 \times 10^{-9}$ (middle), and $2$ (right) respectively, created using the \texttt{Compositional} package \citep{Tsagris2019Compositional}. Below each coutour plot, an instance of the generated contingency table for the ordinal variable $X_i$ and the corresponding discretization of the Gaussian variable $Y_i$ are shown.}
\label{fig:dirichlet}
\end{figure}

In symmetric Dirichlet distributions $\dir(L_i, \nu)$, the parameter $\nu$ controls the density concentration in the $(L_i - 1)$-dimensional probability simplex, which is illustrated in \fref{fig:dirichlet}. By simulating from $\dir(L_i, \nu)$ with $L_i$ fixed and $\nu < 1$, it is more likely to obtain a contingency table where a few cells have much higher frequencies than the others. When most mass concentrates on one end of the table, recovering the original DAGs becomes more difficult, as most correlation information is lost. If $\nu = 1$, it means that we sample uniformly from the simplex. Lastly, if $\nu > 1$, the cell probabilities will be more evenly distributed. 

We consider four expected values $\{2,3,4,5\}$ for $L_i$ (write $\mathbb{E}[L_i]$, $i = 1,\dots,n$) and three values $\{0.8,1,2\}$ for $\nu$. Specifically, we assume that each $L_i$ follows independently a discrete uniform distribution on the interval $[2, 2\mathbb{E}[L_i] - 2]$. When $\mathbb{E}[L_i] = 2$, all variables are binary.

\begin{figure}[H]
\centering
\includegraphics[width = \textwidth]{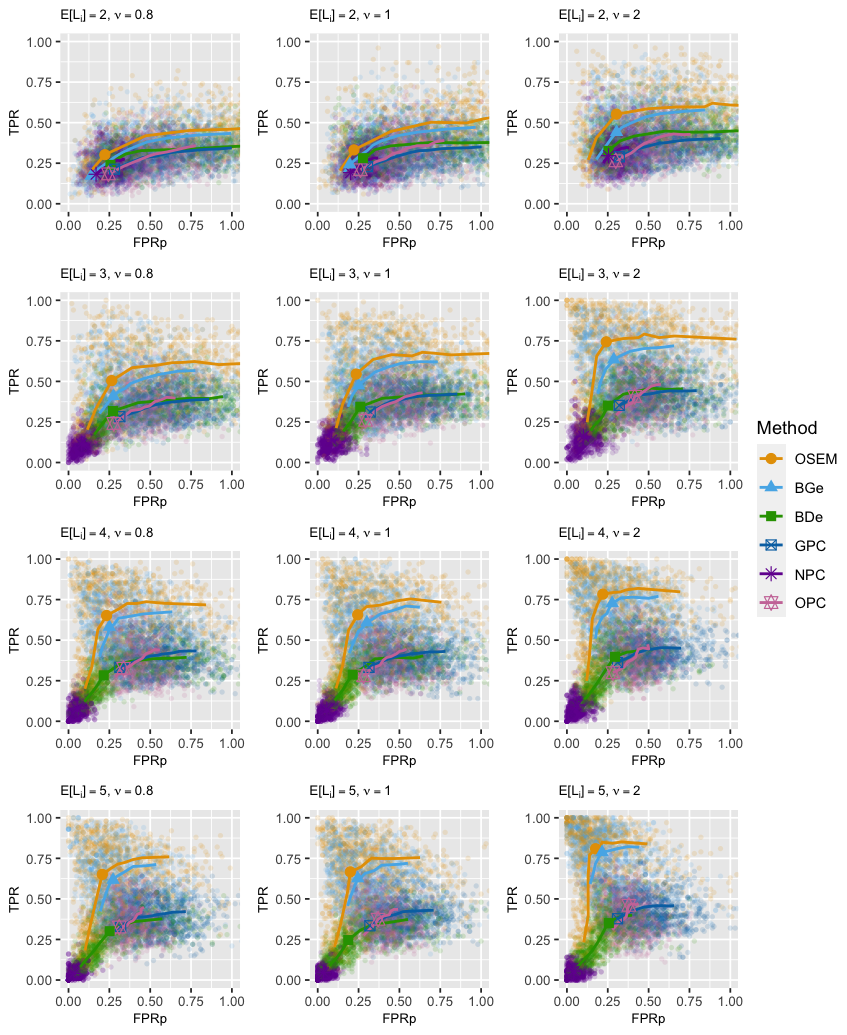}
\caption[Effect of thresholds on the algorithm performance]{The same plot as Figure~\ref{fig:size_test} but with sample size $N=500$, network size $n=20$, expected number of ordinal levels $\mathbb{E}[L_i] \in \{2,3,4,5\}$, and Dirichlet concentration parameter $\nu \in \{0.8,1,2\}$.}
\label{fig:thresholds}
\end{figure}

The ROC curves in \fref{fig:thresholds} again show that our proposed method is superior to all alternatives in every case. With $\nu$ fixed, increasing $\mathbb{E}[L_i]$ significantly improves the performance of both the OSEM algorithm and the method with the BGe score. This is because the resolution of the contingency tables becomes higher. When the discretization is thin enough, the ordinal data should mimic the continuous data on a different scale, and we expect these two curves to move towards the upper left corner of the plot. For similar reasons, the GPC algorithm also slightly improves. Moreover, the rise of the curves gradually becomes slower, which suggests that having more levels is preferable, but too many levels are not necessary for successful structure learning. On the contrary, the performance of the method with the BDe score seems to worsen as $\mathbb{E}[L_i]$ increases, possibly due to the problem of overparameterization. This can be another reason why we should not apply the nominal method on ordinal data.

Conversely, with $\mathbb{E}[L_i]$ fixed, all hybrid methods show various degrees of improvement as $\nu$ increases. The OSEM and the BGe enhance the most, because the contingency tables to be learned from are more well-shaped. The results are least reliable when all variables are binary ($\mathbb{E}[L_i] = 2$) and the cell probabilities are skewed ($\nu = 0.8$). If this is the case, it is however unfortunate that none of the alternative methods will work either. Therefore, the number of thresholds and their positions do play an important role in the recovery of DAG structures, and we should not underestimate the importance of the quality of the data. 

\subsection{Effect of Network Sparsity}

\begin{figure}[H]
\centering
\includegraphics[scale = 0.4]{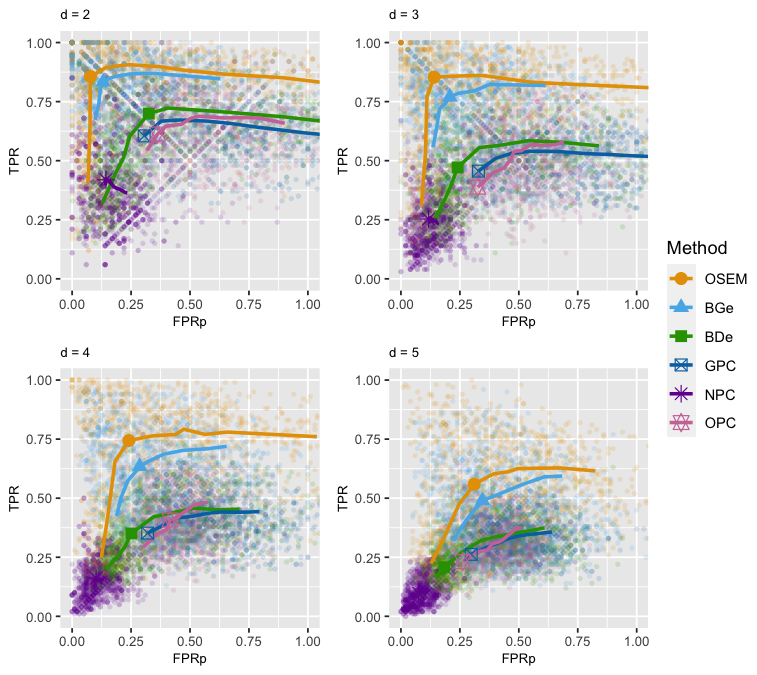}
\caption[Effect of network sparsity]{The same plot as Figure~\ref{fig:size_test} but with sample size $N=500$, network size $n=20$, expected number of ordinal levels $\mathbb{E}[L_i] = 3$, Dirichlet concentration parameter $\nu = 2$, and expected number of neighbours per node $d \in \{2,3,4,5\}$.}
\label{fig:sparsity}
\end{figure}

With network size $n = 20$, sample size $N = 500$, expected number of ordinal levels $\mathbb{E}[L_i] = 3$ and Dirichlet concentration parameter $\nu = 2$, we also look at the effect of network sparsity on the model performance by varying the expected number of neighbours $d$ per node. In \fref{fig:sparsity}, it is clear that the performances of all methods deteriorate as the networks become denser. When $d = 2$, the generated networks are relatively sparse, even the GPC, the OPC and the method with the BDe score can achieve an average TPR of almost $75\%$, and the gap between the OSEM and the method with the BGe score is small. When $d = 5$, however, the average TPRs of all methods drop below $75\%$. In all cases, the OSEM algorithm outperforms the alternatives. In general, structure learning for very dense Bayesian networks is a difficult problem.

\subsection{Runtimes}

Due to its iterative nature, the OSEM algorithm is expected to be slower than other methods. To examine its computational cost, we perform a runtime comparison for DAGs with 12, 20, 30 nodes and 500 data points. The expected number of levels is again 3. The runtimes are associated with the cutoff values at the highest sum of average TPR and $(1 - \text{FPRp})$.

\begin{figure}[H]
\centering
\includegraphics[width = \textwidth]{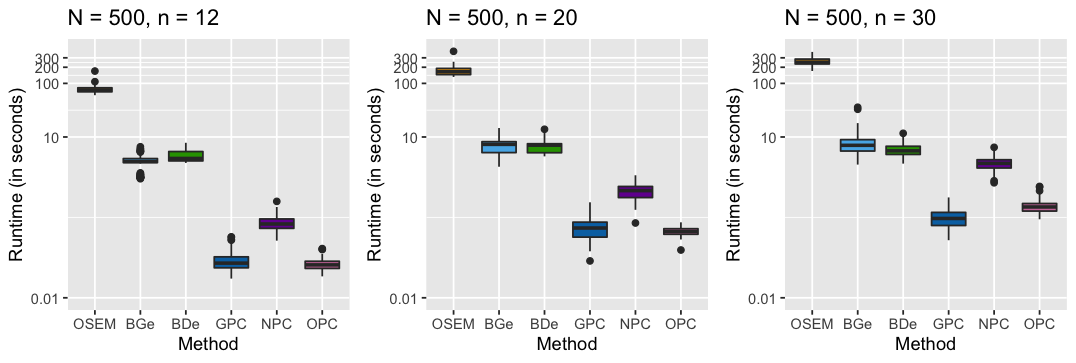}
\vspace{-24pt}
\caption[Runtime comparison]{Runtime comparison in recovering the true patterns for random DAGs with 12, 20, 30 nodes and 500 data points. The runtimes are displayed in seconds, and the y-axis is transformed with $\log_{10}$.}
\vspace{-12pt}
\label{fig:runtime}
\end{figure}

Notice that the runtimes are highly dependent on the algorithmic implementation and the software packages chosen (see \aref{sec:r_func}). For hybrid approaches, the speed and accuracy of the search schemes dominate the overall computational complexity. The runtimes for constraint-based methods rely on how the tests are coded. Therefore, the comparison in \fref{fig:runtime} is intended only as a general reference limited to the functions we chose.

For completeness we present runtimes for the NPC, OPC and GPC algorithms. They are less interesting due to their unsatisfying performance in the recovery of DAGs. On average, both hybrid methods with the BDe and BGe scores take less than $10$ seconds to finish for all three network sizes. Our OSEM algorithm, however, takes around $1.5$, $2.5$ and $3.5$ minutes respectively in order to converge, which is sometimes more than $20$ times slower than a pure hybrid approach without EM iterations. Such a loss in computational efficiency is mainly due to sampling from the $n$-dimensional truncated multivariate normal distributions, especially when $n$ is large. Consequently, the current version of our method is only feasible for small-to-medium networks. For larger networks, one may wish to conduct a pre-selection of the variables before using such models for inference. 

\subsection{Comparison with Existing Mixed Methods}
\label{sec:mixed}

\begin{figure}[t]
\centering
\includegraphics[width = \textwidth]{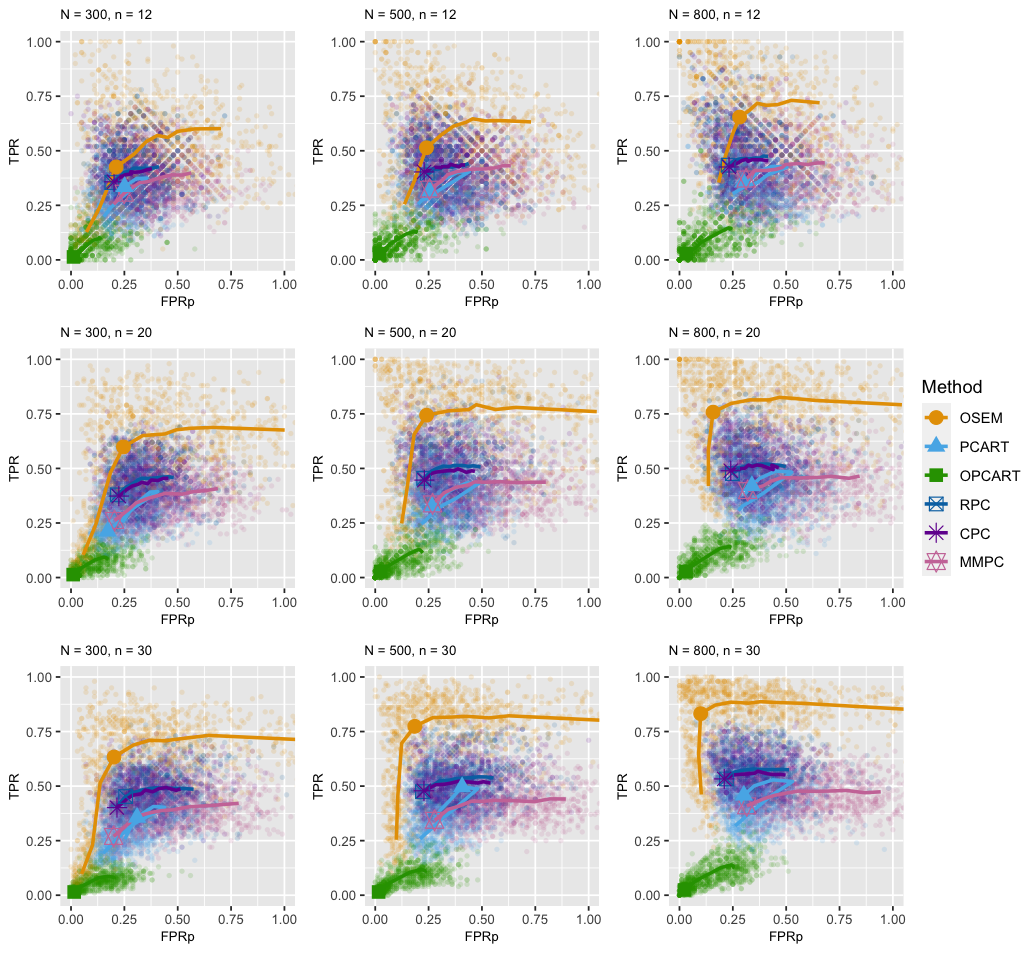}
\vspace{-24pt}
\caption[Comparison with existing mixed methods]{The comparison of the performance in recovering the true pattern between our OSEM algorithm and five other approaches: PCART, OPCART, CPC, RPC, and MMPC. The ROC curves are created as in \fref{fig:size_test}.}
\label{fig:sizes_mixed}
\end{figure}

We additionally benchmark the OSEM algorithm against the most recent structure learning methods for mixed (continuous, nominal and ordinal) variables. If we consider only the ordinal data, our OSEM algorithm clearly outperforms the following approaches in simulations (\fref{fig:sizes_mixed}):
\begin{itemize}
\itemsep0em
    \item The copula PC algorithm of \cite{cui2016copula} (CPC); To be consistent with the paper mentioned, we also include the rank PC algorithm of \cite{harris2013pc} (RPC);
    \item PC algorithm with a symmetric test based on two asymmetric likelihood-ratio tests \citep{tsagris2018constraint} (MMPC);
    \item Two score-based approaches based on classification and regression trees \citep{talvitie2019learning}: PCART and OPCART. OPCART is an extension of PCART to accommodate ordinal response variables in decision trees. Note that OPCART performs much worse than PCART. With the segmentation prior parameter $\kappa$ set to $\frac{1}{4}$, as suggested by the authors, the estimated OPCART graph is always very sparse no matter how one changes the other hyperparameters. If one increases $\kappa$, then OPCART has similar performance as PCART but requires much more computational resources due to the segmentation step, which takes around an hour to learn one network with $20$ nodes. In the section below, we also show that OPCART has worse predictive performance than PCART.
\end{itemize}

All implementations are summarised in \aref{sec:r_func}.

\subsection{More about Predictive Performance}
\label{sec:pred_supp}

In \tref{tab:real_data sets} one can find all five data sets used to assess the predictive performance of the OSEM algorithm. In the pre-processing step, we have categorized two numerical variables in the Contraceptive Method Choice data set: \textit{wife's age} ($0$: $[0, 20)$; $1$: $[20,30)$; $2$: $[30,40)$; $3$: $[40,\infty)$) and \textit{number of children ever born} ($0$: $[0,1]$; $1$: $[2,3]$; $2$: $[4,5]$, $3$: $[6,\infty)$). We have also removed the \textit{class} variable with $22$ levels from the Primary Tumor data set.

\begin{figure}[H]
\centering
\includegraphics[width = \textwidth]{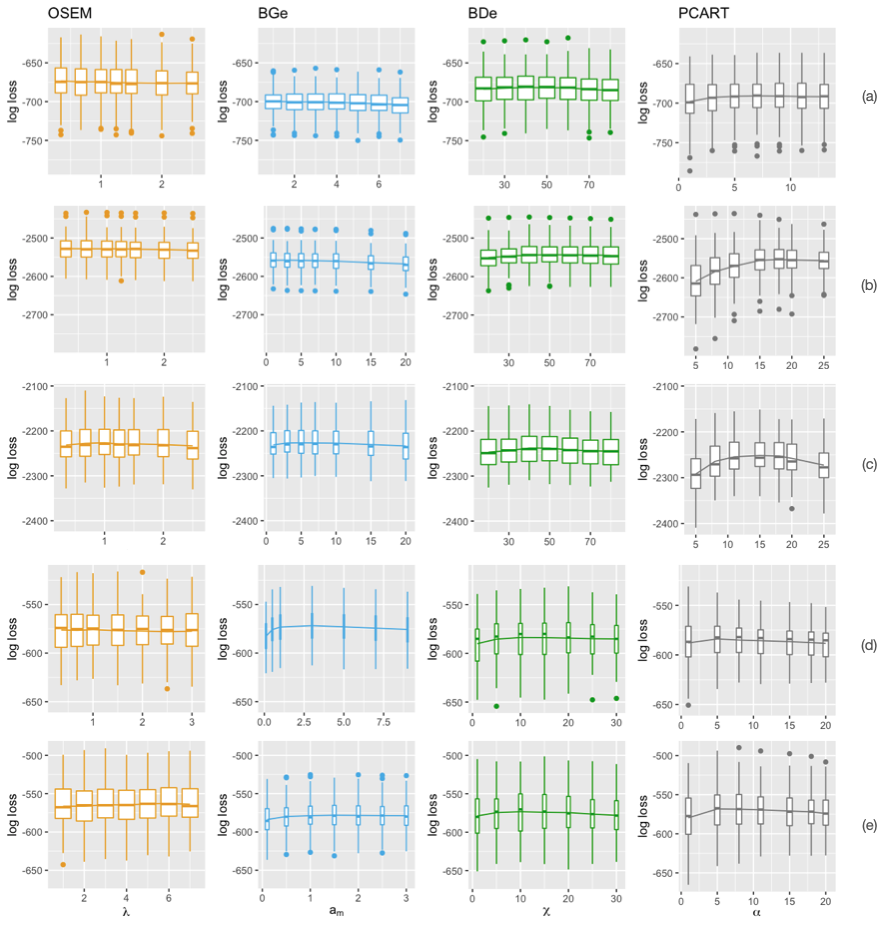}
\vspace{-24pt}
\caption{The comparison of log loss for four methods (OSEM, BGe, BDe, PCART) and five data sets: (a) Congressional Voting Records, (b) Contraceptive Method Choice, (c) OCD and Depression, (d) Primary Tumor, and (e) SPECT Heart.}
\label{fig:log_loss_all}
\end{figure}

\begin{figure}[H]
\centering
\includegraphics[width = 0.7\textwidth]{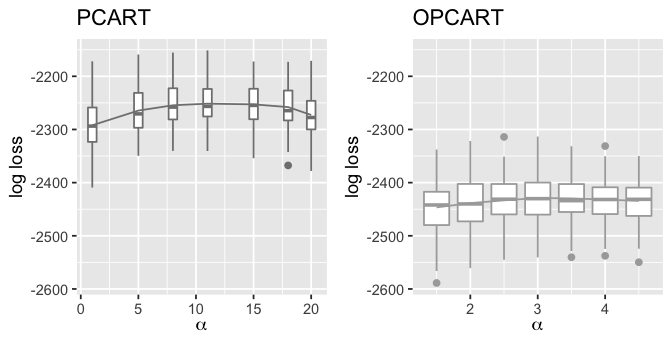}
\vspace{-12pt}
\caption{The comparison of log loss between PCART and OPCART for the OCD and Depression data set.}
\label{fig:log_loss_opcart}
\end{figure}

\begin{figure}[H]
\centering
\includegraphics[width = \textwidth]{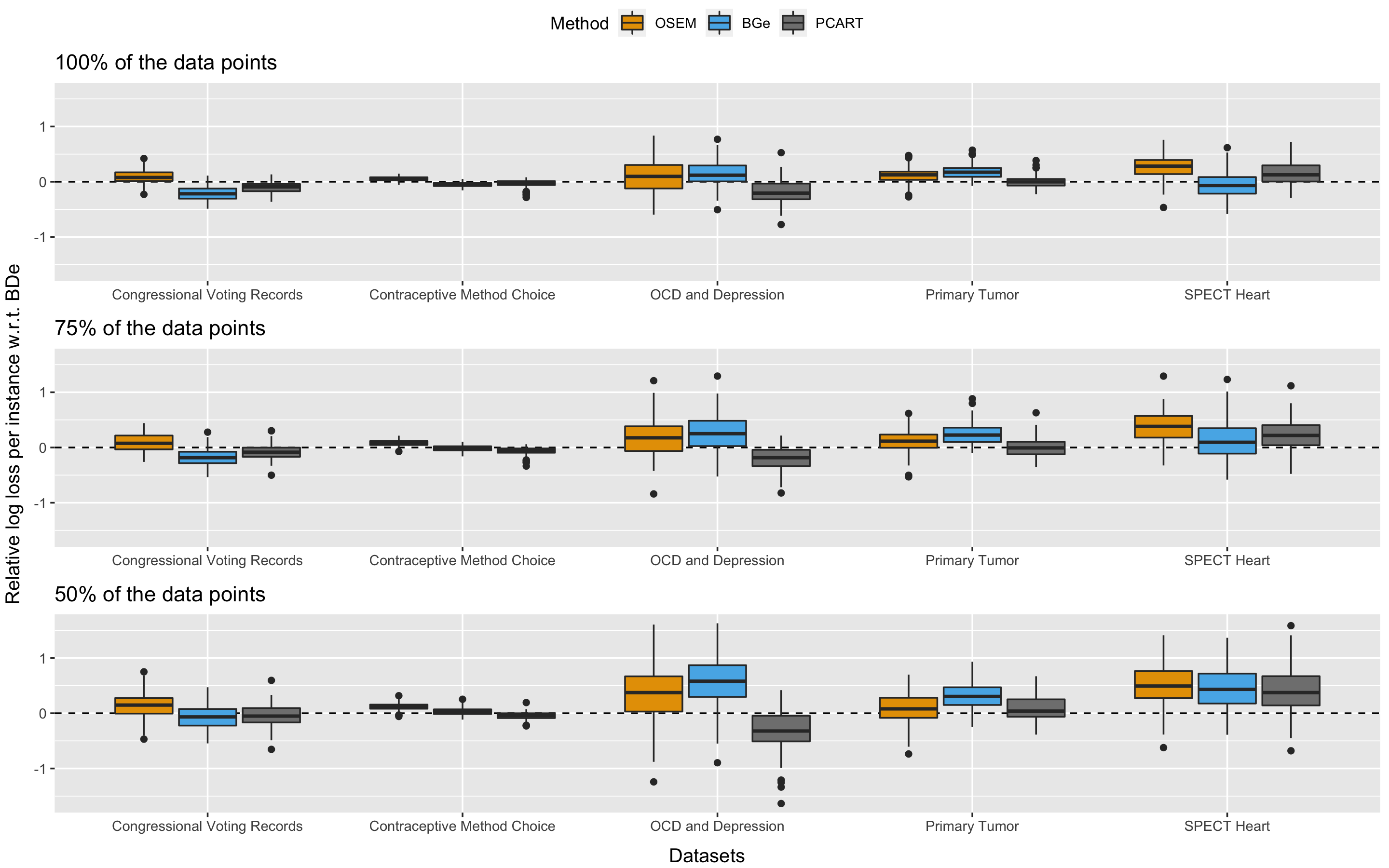}
\vspace{-12pt}
\caption{The comparison of log loss with decreasing sample sizes. In each simulation run, we first randomly subsample a proportion of the data points ($75\%$ or $50\%$) from the data sets, followed by structure learning and log loss computation.}
\label{fig:log_loss_subsample}
\end{figure}

In cross validation, we first randomly split each data set into $80\%$ training data and $20\%$ test data. Using the training data, we find the optimal model with respect to each of the methods being compared (BDe, BGe, OSEM, PCART/OPCART). Conditioned on the estimated model $\hat{\mathcal{M}}$, we evaluate the discrete (negative) log loss (log-likelihood) function on the test data,
\begin{equation}
    \log P(\mathbf{X}_{\text{test}} \mid \hat{\mathcal{M}}) =  \sum^{N_{\text{test}}}_{i = 1}\sum^L_{l = 1} \mathds{1}_{\mathbf{x}_{il}} \log \hat{p}_{l},
\end{equation}
where $N_{\text{test}}$ is the number of unseen test cases, $L$ is the total number of combinations of all ordinal levels, $\mathds{1}_{\mathbf{x}_{il}}$ indicates whether the $i^{th}$ data point matches the $l^{th}$ combination, and $\hat{p}_{l}$ is the probability of observing the $l^{th}$ combination given the model. In particular, we select $\hat{p}_{l}$ from the probability tables computed for each method being compared:
\begin{itemize}
    \item BDe: from the BDe posterior predictive distribution \citep{heckerman1995learning};
    \item PCART/OPCART: from the BDe posterior predictive distribution;
    \item OSEM: by integrating the multivariate normal distribution with zero mean vector and the optimal correlation matrix from the OSEM output over regions defined by the estimated thresholds $\hat{\bm{\alpha}}$ from \eref{eq:threshold};
    \item BGe: by integrating the multivariate normal distribution with zero mean vector and the maximum a posteriori correlation matrix based on the estimated DAG \citep{viinikka2020towards} over regions defined by the estimated thresholds $\hat{\bm{\alpha}}$ from OSEM.
\end{itemize}
For each method, we repeat the above process $100$ times over a range of tuning parameters and obtain \fref{fig:log_loss_all}, where the optimal values are selected and summarised in \fref{fig:log_loss}. We also provide an example in \fref{fig:log_loss_opcart} to show that OPCART has worse predictive performance than PCART and hence it is omitted from the main text. In addition, we evaluate the log loss with decreasing number of data points by subsampling the data sets (\fref{fig:log_loss_subsample}). The relative log loss of the BDe approach slightly worsens in comparison to the other three methods, possibly due to overparameterization. As sample size decreases, the uncertainty consistently increases, and the advantage of OSEM over BGe appears less pronounced.

\begin{figure}[H]
\centering
\includegraphics[width = \textwidth]{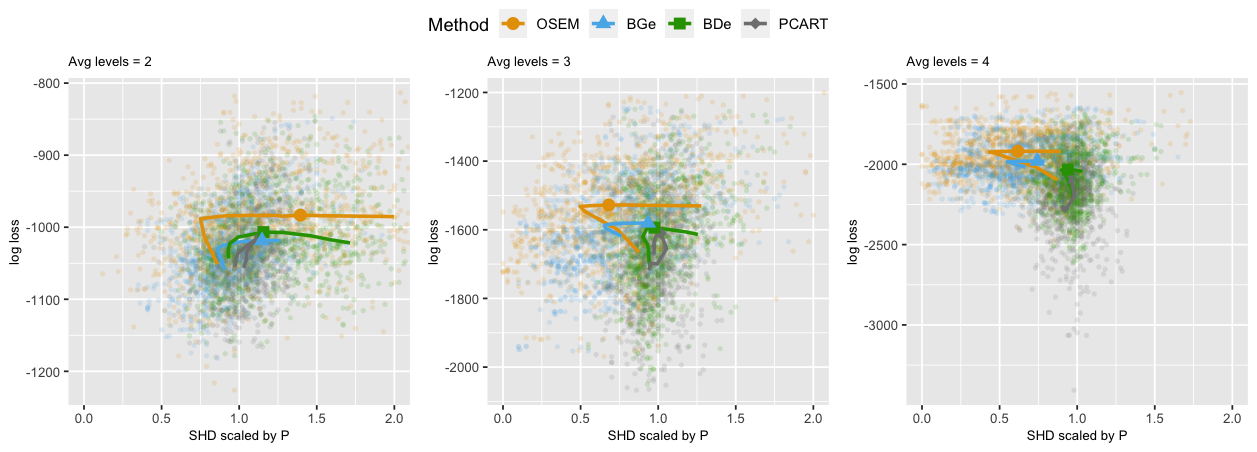}
\vspace{-24pt}
\caption{The comparison of log loss against SHD scaled by P for four methods (OSEM, BGe, BDe, PCART) using synthetic data. The lines are created by interpolating the average log loss against the average SHD scaled by P at each penalization value. The solid points correspond to the optimal/maximum log loss for each method.}
\label{fig:log_loss_SHD}
\end{figure}

As a sanity check, we evaluate the predictive performance using synthetic data sets where we know the ground truth networks. Considering $N = 500$, $n = 20$ and three expected number of levels $\{2,3,4\}$, we plot in \fref{fig:log_loss_SHD} the log loss against the structural hamming distance scaled by P as defined in \sref{sec:metrics}. As we expect, the OSEM algorithm dominates the other three methods in terms of both log loss and SHD. As the average number of levels increases, the performance of BGe becomes closer to OSEM, which is consistent with the observations in \aref{sec:thresholds}. Interestingly, the optimum log loss for OSEM and BGe (solid points) do not match the minimum SHD (the top left corner of the curves), and the log loss is very flat across a wide range of tuning parameters. This may suggest that overfitting in Gaussian-based BNs can be beneficial in terms of predictive power, and the strength of regularization needs to be stronger for optimal structure learning than for prediction.  

\newpage

\subsection{\texttt{R} Implementations}
\label{sec:r_func}

\footnotetext[1]{\url{https://github.com/xgluo/OSEM}}
\footnotetext[2]{\url{https://github.com/cuiruifei/CausalMissingValues}}
\footnotetext[3]{\url{https://github.com/ttalvitie/pcart}}
\footnotetext[4]{\url{https://github.com/xgluo/pcart}. Note that the original \texttt{rpcart} package does not have an \texttt{R} implementation for OPCART.}

\begin{table}[H]
\small
    \centering
    \begin{tabular}{p{0.1\linewidth}|p{0.115\linewidth}|p{0.375\linewidth}|p{0.3\linewidth}}
      \hline
      \textbf{Method} & \textbf{\texttt{R} package} & \textbf{\texttt{R} functions} & \textbf{Reference} \\
      \hline
      NPC & \texttt{pcalg} & \texttt{pc}, \texttt{disCItest} & \cite{neapolitan2004learning, kalisch2012causal} \\
      OPC & \texttt{bnlearn} & \texttt{pc.stable(..., test = "jt")} & \cite{scutari2009learning, musella2013pc}\\
      GPC & \texttt{pcalg} & \texttt{pc}, \texttt{gaussCItest} & \cite{kalisch2012causal} \\
      BDe & \texttt{BiDAG} & \texttt{scoreparameters("bdecat",...)}, \texttt{iterativeMCMC} & \cite{heckerman1995learning, suter2021bayesian}\\
      BGe & \texttt{BiDAG} & \texttt{scoreparameters("bge",...)}, \texttt{iterativeMCMC} & \cite{heckerman1995learning, suter2021bayesian}\\
      OSEM & \texttt{BiDAG} & \texttt{iterativeMCMC}, OSEM\footnotemark & \cite{suter2021bayesian}\\
      \hline
      RPC & \texttt{pcalg} & \texttt{pc}, \texttt{gaussCItest}, \texttt{cor(..., method = "kendall")} & \cite{kalisch2012causal,harris2013pc} \\
      CPC & \texttt{pcalg} & \texttt{pc}, \texttt{gaussCItest}, \texttt{inferCopulaModel}\footnotemark & \cite{kalisch2012causal,cui2016copula} \\
      MMPC & \texttt{MXM} & \texttt{pc.skel(..., method = "comb.mm")}, \texttt{pc.or} & \cite{tsagris2018constraint} \\
      PCART & \texttt{rpcart}\footnotemark, \texttt{BiDAG} & \texttt{opt.pcart.cat}, \texttt{iterativeMCMC} & \cite{talvitie2019learning, suter2021bayesian}
       \\
      OPCART & \texttt{rpcart}\footnotemark, \texttt{BiDAG} & \texttt{opt.pcart}, \texttt{iterativeMCMC} & \cite{talvitie2019learning, suter2021bayesian}
    \end{tabular}
    \caption{\texttt{R} implementations for simulation studies. The first six methods are discussed in \sref{sec:experiments}, and the last five methods can be found in \aref{sec:mixed}.}
    \label{tab:R_func}
\end{table}

\section{Additional Table and Figures}

\begin{table}[H]
    \centering
    \begin{tabular}{l | l || l | l}
      \hline
      \multicolumn{4}{c}{\textbf{Symptoms and the associated abbreviations}} \\
      \hline
      \multicolumn{2}{c||}{Y-BOCS-SR} & \multicolumn{2}{c}{QIDS-SR} \\
      \hline
      \hline
      Time consumed by obsessions & \textit{obtime} & Sleep-onset insomnia & \textit{onset} \\
      Interference due to obsessions & \textit{obinterfer} & Middle insomnia & \textit{middle} \\
      Distress caused by obsessions & \textit{obdistress}  & Early morning awakening & \textit{late}  \\
      Difficulty resisting obsessions & \textit{obresist}  & Hypersomnia & \textit{hypersom}  \\
      Difficulty controlling obsessions & \textit{obcontrol}  &  Sadness & \textit{sad}  \\
      Time consumed by compulsions & \textit{comptime}  & Decreased/Increased appetite & \textit{appetite}  \\
      Interference due to compulsions & \textit{compinterf}  & Weight loss/gain & \textit{weight}  \\
      Distress caused by compulsions & \textit{compdis}  & Concentration impairment & \textit{concen}  \\
      Difficulty resisting compulsions & \textit{compresist}  & Guilt and self-blame & \textit{guilt}  \\
      Difficulty controlling compulsions & \textit{compcont}  & Suicidal thoughts or attempts & \textit{suicide}  \\
       & & Anhedonia & \textit{anhedonia}  \\
       & & Fatigability & \textit{fatigue}  \\
       & & Psychomotor slowing & \textit{retard}  \\
       & & Psychomotor agitation & \textit{agitation}  \\
      \hline
    \end{tabular}
    \caption{Symptoms and the associated abbreviations on the self report versions of the Yale-Brown Obsessive-Compulsive Scale (Y-BOCS-SR) \citep{steketee1996yale} and the Quick Inventory of Depressive Symptomatology (QIDS-SR) \citep{rush200316}. We combine the symptoms ``Increased appetite'' and ``Decreased appetite'' into one variable \textit{appetite} with $7$ levels. Analogously, the symptoms ``weight gain'' and ``weight loss'' are combined into the variable \textit{weight}.}
    \label{tab:OCDRogers}
\end{table}

\newpage

\begin{figure}[H]
\centering
\includegraphics[width = 0.9\textwidth]{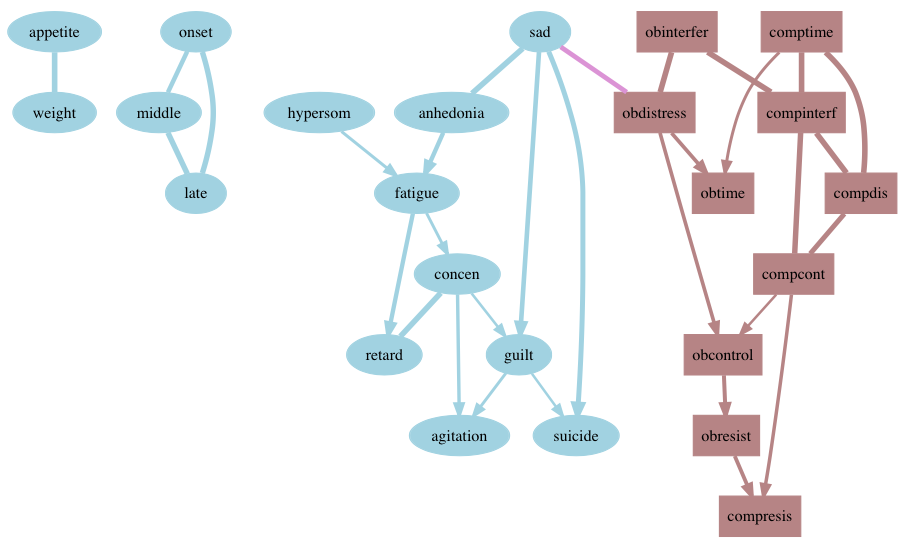}
\caption{CPDAG of the reproduced Bayesian network using the method described in \citep{mcnally2017co}. As in \fref{fig:psych_OSEM}, rectangles represent nodes related to OCD symptoms, ellipses represent nodes related to depression symptoms, and edges between nodes with different colors are highlighted with pink.}
\label{fig:psych_mcnally}
\end{figure}

\begin{figure}[H]
\centering
\includegraphics[width = \textwidth]{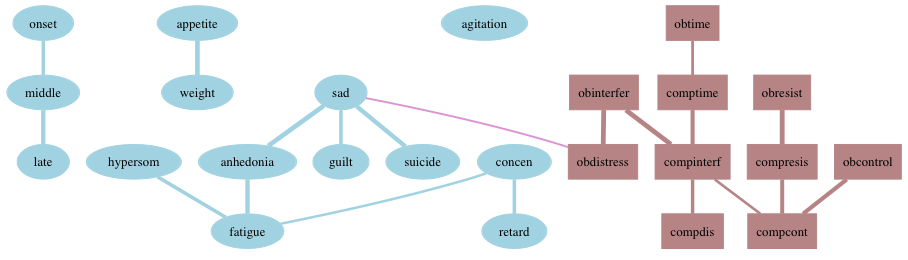}
\caption{CPDAG estimated via the hybrid method, as described in \sref{sec:experiments}, with the BDe score and the nominal PC output as the initial search space. The penalty for the BDe score $\chi$ is chosen to be $1.5$. As in \fref{fig:psych_OSEM}, rectangles represent nodes related to OCD symptoms, ellipses represent nodes related to depression symptoms, and edges between nodes with different colors are highlighted with pink.}
\label{fig:psych_BDe}
\end{figure}

\begin{figure}[H]
\centering
\includegraphics[scale = 0.55]{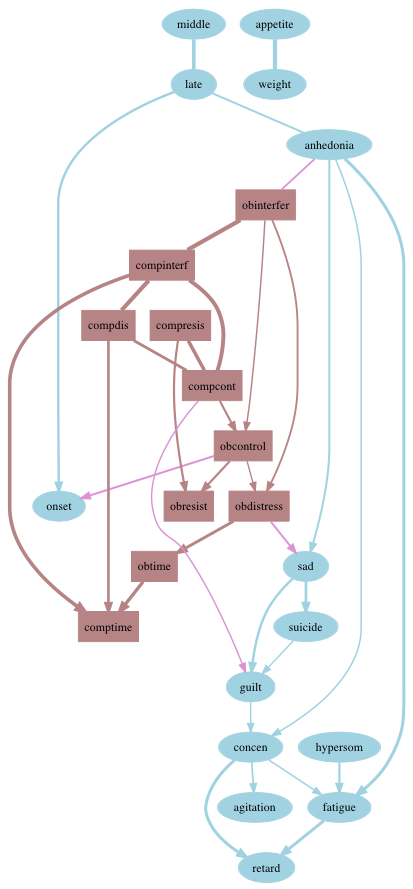}
\caption{CPDAG estimated via the hybrid method, as described in \sref{sec:experiments}, with the BGe score and the Gaussian PC output as the initial search space. The penalty for the BGe score $a_m$ is chosen to be $0.05$. As in \fref{fig:psych_OSEM}, rectangles represent nodes related to OCD symptoms, ellipses represent nodes related to depression symptoms, and edges between nodes with different colors are highlighted with pink.}
\label{fig:psych_BGe}
\end{figure}

\begin{figure}[H]
\centering
\includegraphics[width = \textwidth]{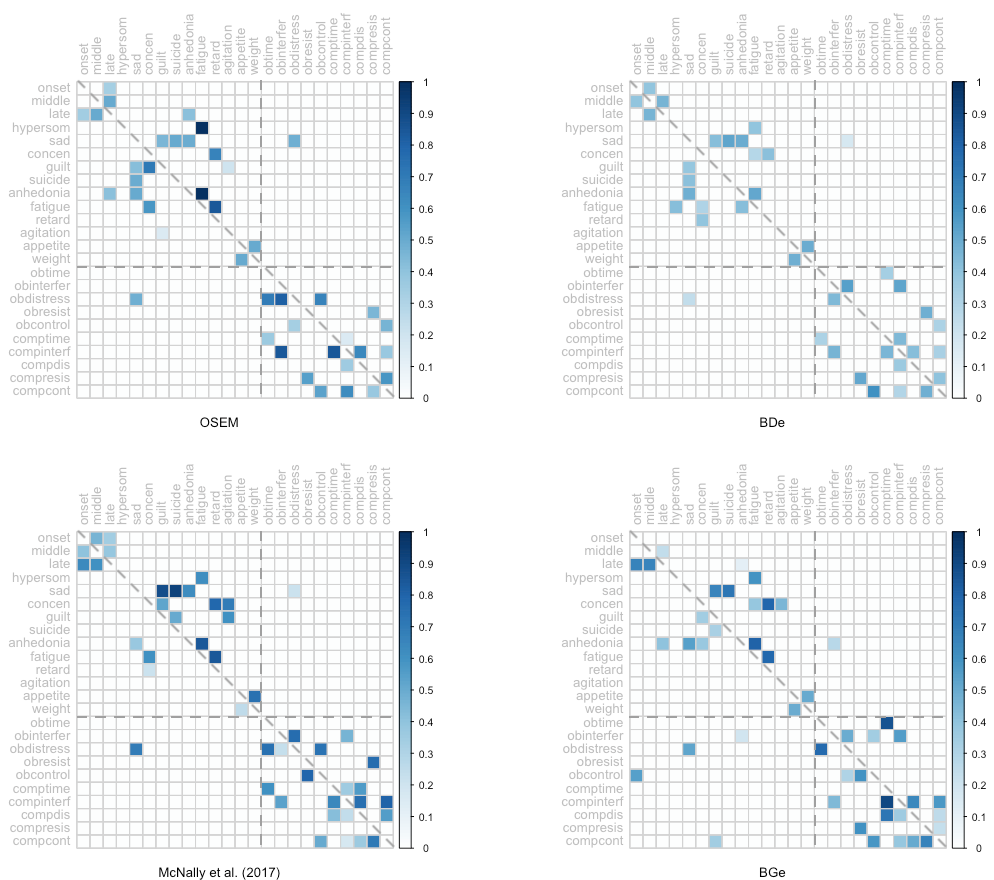}
\caption{Heatmaps for the CPDAG adjacency matrices estimated using the four methods mentioned in \sref{sec:psych}. The shade in the grid represents the percentage of time a directed edge appears in the $500$ Bootstrapped CPDAGs, where an undirected edge counts half for each direction. The darker the shade, the more frequent the corresponding directed edge occurs.}
\label{fig:heatmap}
\end{figure}

\clearpage


\bibliography{20-1338}

\end{document}